\headline={\ifnum\pageno=1\hfil\else\hfil\tenrm--\ \folio\ --\hfil\fi}
\footline={\hfil}
\hsize=6.0truein
\vsize=8.54truein
\hoffset=0.25truein
\voffset=0.25truein
\baselineskip=18pt
%
%
\tolerance=9000
\hyphenpenalty=10000
%
%
%

\font\mbf=cmmib10 \font\mbfs=cmmib10 scaled 833
\font\msybf=cmbsy10 \font\msybfs=cmbsy10 scaled 833

%
%
%

\textfont9=\mbf \scriptfont9=\mbfs \scriptscriptfont9=\mbfs

\textfont10=\msybf \scriptfont10=\msybfs \scriptscriptfont10=\msybfs
%
%
\mathchardef\alpha="710B
\mathchardef\beta="710C
\mathchardef\gamma="710D
\mathchardef\delta="710E
\mathchardef\epsilon="710F
\mathchardef\zeta="7110
\mathchardef\eta="7111
\mathchardef\theta="7112
\mathchardef\iota="7113
\mathchardef\kappa="7114
\mathchardef\lambda="7115
\mathchardef\mu="7116
\mathchardef\nu="7117
\mathchardef\xi="7118
\mathchardef\pi="7119
\mathchardef\rho="711A
\mathchardef\sigma="711B
\mathchardef\tau="711C
\mathchardef\upsilon="711D
\mathchardef\phi="711E
\mathchardef\chi="711F
\mathchardef\psi="7120
\mathchardef\omega="7121
\mathchardef\varepsilon="7122
\mathchardef\vartheta="7123
\mathchardef\varpi="7124
\mathchardef\varrho="7125
\mathchardef\varsigma="7126
\mathchardef\varphi="7127
\mathchardef\nabla="7272
\mathchardef\cdot="7201
%
%
%

%
%

%
%

%
%

%
%
\newcount\eqnumber
\eqnumber=1
%
\def\new{{\the\eqnumber}\global\advance\eqnumber by 1}
%
%
\def\ref#1{\advance\eqnumber by -#1 \the\eqnumber
     \advance\eqnumber by #1 }
%
%
\def\last{\advance\eqnumber by -1 {\the\eqnumber}\advance 
     \eqnumber by 1}
%
%
\def\eqnam#1{\xdef#1{\the\eqnumber}}
%
%
%
\def\refindent{\par\noindent\hangindent=3pc\hangafter=1 }
\def\aa#1#2#3{\refindent#1, A\&A, #2, #3}

\def\apj#1#2#3{\refindent#1, {\it Ap. J.}, {\bf#2}, #3.}

\def\mnras#1#2#3{\refindent#1, {\it M.N.R.A.S.}, {\bf#2}, #3.}

%
\def\sect#1 {
  \bigbreak
  \centerline{\bf #1}
  \bigskip}
\def\subsec#1#2 {
  \bigbreak
  \centerline{#1.~{\it #2}}
  \bigskip}
\newcount\figno
\figno=0
\def\figure{\global\advance\figno by 1 Figure~\the\figno.~}
%
%

\def\>{$>$}
\def\<{$<$}

\def\simlt{\lower.5ex\hbox{$\; \buildrel < \over \sim \;$}}
\def\simgt{\lower.5ex\hbox{$\; \buildrel > \over \sim \;$}}
\def\sqr#1#2{{\vcenter{\hrule height.#2pt
      \hbox{\vrule width.#2pt height#1pt \kern#1pt
         \vrule width.#2pt}
      \hrule height.#2pt}}}

%
%

\def\today{\ifcase\month\or
	January\or February\or March\or April\or May\or June\or
	July\or August\or Setrueptember\or October\or November\or December\fi
	\space\number\day, \number\year}
%
\def\head#1{\headline={\ifnum\pageno>1
	{\tenrm #1} \hfil Page \folio
	\else\hfil\fi}}
%
\def\ref #1;#2;#3;#4{\par\pp #1, {\it #2}, {\bf #3}, #4}
\def\book #1;#2;#3{\par\pp #1, {\it #2}, #3}
\def\rep #1;#2;#3{\par\pp #1, #2, #3}
%
%

\newbox\grsign \setbox\grsign=\hbox{$>$}
\newdimen\grdimen \grdimen=\ht\grsign
\newbox\laxbox \newbox\gaxbox
\setbox\gaxbox=\hbox{\raise.5ex\hbox{$>$}\llap
     {\lower.5ex\hbox{$\sim$}}}\ht1=\grdimen\dp1=0pt
\setbox\laxbox=\hbox{\raise.5ex\hbox{$<$}\llap
     {\lower.5ex\hbox{$\sim$}}}\ht2=\grdimen\dp2=0pt


\vsize 8.7in

\def\doublespace{\baselineskip 22.76 pt}

\hsize 6.5 true in
\hoffset=0. true in
\voffset= 0.6 true in
\def\mathnew{\mathsurround=0pt}
\def\simov#1#2{\lower .5pt\vbox{\baselineskip0pt \lineskip-.5pt
\ialign{$\mathnew#1\hfil##\hfil$\crcr#2\crcr\sim\crcr}}}

\font\ninerm=cmr7 scaled 1200
\font\sevenrm=cmr5 scaled 1200
\font\twelvei=cmmi10 scaled 1200
\font\ninei=cmmi7 scaled 1200
\font\seveni=cmmi5 scaled 1200
\font\twelvesy=cmsy10 scaled 1200
\font\ninesy=cmsy7 scaled 1200
\font\sevensy=cmsy5 scaled 1200

\font\ninebf=cmbx7 scaled 1200
\font\sevenbf=cmbx5 scaled 1200

\skewchar\twelvei='177 \skewchar\ninei='177 \skewchar\seveni='177
\skewchar\twelvesy='60 \skewchar\ninesy='60 \skewchar\sevensy='60
\font\elevenrm=cmr10 scaled 1100
\font\ninerm=cmr7 scaled 1100
\font\sevenrm=cmr5 scaled 1100
\font\eleveni=cmmi10 scaled 1100
\font\ninei=cmmi7 scaled 1100
\font\seveni=cmmi5 scaled 1100
\font\elevensy=cmsy10 scaled 1100
\font\ninesy=cmsy7 scaled 1100
\font\sevensy=cmsy5 scaled 1100
\font\elevenex=cmex10 scaled 1100
\font\elevenbf=cmbx10 scaled 1100
\font\ninebf=cmbx7 scaled 1100
\font\sevenbf=cmbx5 scaled 1100
\font\elevenit=cmti10 scaled 1100
\font\elevensl=cmsl10 scaled 1100
\font\eleventt=cmtt10 scaled 1100
\skewchar\eleveni='177 \skewchar\ninei='177 \skewchar\seveni='177
\skewchar\elevensy='60 \skewchar\ninesy='60 \skewchar\sevensy='60
\def\elevenpoint{\def\rm{\fam0 \elevenrm}
  \textfont0=\elevenrm \scriptfont0=\ninerm \scriptscriptfont0=\sevenrm
  \rm
  \textfont1=\eleveni \scriptfont1=\ninei \scriptscriptfont1=\seveni
  \def\mit{\fam1 } \def\oldstyle{\fam1 \eleveni}
  \textfont2=\elevensy \scriptfont2=\ninesy \scriptscriptfont2=\sevensy
  \def\cal{\fam2 }
  \textfont3=\elevenex \scriptfont3=\elevenex \scriptscriptfont3=\elevenex
  \textfont\itfam=\elevenit \def\it{\fam\itfam\elevenit}
  \textfont\slfam=\elevensl \def\sl{\fam\slfam\elevensl}
  \textfont\bffam=\elevenbf \scriptfont\bffam=\ninebf
    \scriptscriptfont\bffam=\sevenbf \def\bf{\fam\bffam\elevenbf}
  \textfont\ttfam=\eleventt \def\tt{\fam\ttfam\eleventt}
  }
\def\folio{\ifnum\pageno=1\nopagenumbers\else\number\pageno\fi}
\font\simlessrgebf=cmbx10 scaled 1600
\font\twelvess=cmss10 scaled 1200
\elevenpoint
\baselineskip 15pt
\def\ref{\par\noindent\hangindent=2pc \hangafter=1 }

\def\physrev{{\it Phys. Rev.}}
\def\apj{{\it Ap.~J.}}

\def\mnras{{\it M.N.R.A.S.}}

\def\integral{\int\limits_{1}^{\infty}}
\def\kyl{\ln\Lambda}

\null
\centerline{Submitted to the Editor of the Astrophysical Journal}
\centerline{Revised April 22, 1997}
\null\vskip 0.85 true in
\centerline{\simlessrgebf Self-Consistent Fokker-Planck Treatment}
\vskip 0.2in
\centerline{\simlessrgebf Of Particle Distributions in Astrophysical Plasmas}
\vskip 1.7in
\centerline{{\bf Sergei Nayakshin}$^1$ and {{\bf Fulvio 
Melia}$^{1,2,}$}\footnote{\hbox{$^{\dag}$}}{Presidential Young Investigator}}
\vskip 0.05in
\centerline{\sl $^1$Department of Physics, University of Arizona, Tucson, AZ
 85721}
\centerline{\sl [serg@physics.arizona.edu]}
\centerline{\sl $^2$Department of Astronomy, University of Arizona, Tucson, AZ 85721}
\centerline{\sl [melia@as.arizona.edu]}
\vfill\eject
\null
\centerline{\bf ABSTRACT}
\bigskip
{\twelvess
High-energy, multi-component plasmas in which pair creation and
annihilation, lepton-lepton scattering, lepton-proton scattering,
and Comptonization all contribute to establishing the particle
and photon distributions, are present in a broad range of
compact astrophysical objects.  The different constituents are
often not in equilibrium with each other, and this mixture of 
interacting particles and radiation can produce substantial
deviations from a Maxwellian profile for the lepton distributions.
Earlier work has included much of the microphysics needed to account 
for electron-photon and electron-proton interactions,
but little has been done to handle the redistribution of the particles 
as a result of their Coulomb interaction with themselves. The most 
detailed analysis thus far for finding the exact electron distribution 
appears to have been done within the framework of non-thermal
models, where the electron distribution is approximated as a thermal one
at low energy with a non-thermal tail at higher energy. 
Recent attention, however, has been focused on thermal models.

Our goal here is to
use a Fokker-Planck approach in order to develop a fully self-consistent 
theory for the interaction of arbitrarily distributed particles and 
radiation to arrive at an accurate representation of the 
high-energy plasma in these sources. We conduct several tests representative
of two dominant segments of parameter space. For high source 
compactness of the total radiation field, $l\sim 10^2$, we find that
although the electron distribution deviates substantially from a Maxwellian,
the resulting photon spectra are insensitive to the exact electron shape,
in accordance with some earlier results.  For low source compactness, $l\sim$ few,
and an optical depth $\simlt 0.2$, however, we find that both the electron
distribution and the photon spectra differ strongly from what they
would be in the case of a Maxwellian distribution.
In addition, for all values of compactness, we find that different electron 
distributions lead to different positron number densities and proton 
equilibrium temperatures. This means that the ratio of radiation pressure
to proton pressure is strongly dependent on the lepton distribution,
which might lead to different configurations of hydrostatic equilibrium. 
This, in turn, may change the compactness, optical depth, and 
heating and cooling rates, and therefore lead to an additional
change in the spectrum. An important result of our analysis, is the
derivation of useful, approximate analytical forms for the electron
distribution in the case of strongly non-Maxwellian plasmas.
}
\bigskip\noindent {\it Subject headings}: accretion disks---black hole 
physics---galaxies:  Seyfert---gamma rays: theory---plasmas---radiative transfer
\vfil\eject

\centerline{\bf 1. Introduction}
\medskip
There is a great deal of interest in the physics of relativistic, multi-component,
astrophysical plasmas, which appear to be present in a wide variety of high-energy
emitting objects, such as neutron stars, black holes, and Active Galactic Nuclei (AGNs).
The particle distributions in these sources depend critically on key physical
processes, such as pair creation and annihilation, lepton-lepton scattering,
lepton-proton scattering, and Comptonization.  Often, the particle population and
the radiation field are not in equilibrium, and 
it is possible that in some cases, such as the inner 
coronal regions in black hole systems like Cygnus X-1 and 1E1740.7-2942, 
the electrons and protons are themselves not equilibrated, with
the proton ``temperature'' greatly exceeding that of the electrons and positrons
(Shapiro, Lightman \& Eardley 1976;  for a more recent discussion and more extensive
list of references, see also Misra \& Melia 1996).

The attention paid to these plasmas is fostered by continued, exciting observations
of the underlying objects by a number of high-energy instruments on satellites 
such as the
{\sl Compton Gamma-Ray Observatory (CGRO)}. For example, OSSE and COMPTEL on 
{\sl CGRO} have recently revealed that persistent MeV gamma-rays are emitted by
several Galactic black hole candidates (GBHCs), including Cygnus X-1 (Johnson
et al. 1993;  McConnell et al. 1994) and GRO J0422+32 (van Dijk et al. 1995).
Single component thermal models (e.g., Sunyaev \& Titarchuk 1980;  Titarchuk
1994) have difficulty accounting for these high-energy tails, which strongly
hint at nonthermal processes.  In these systems, not only is a substantial 
fraction of the radiation produced externally and therefore not initially
in thermal equilibrium with the electron-positron pair plasma, but
the combination of pair creation/annihilation (Liang 1979; Kusunose 1987; 
Svensson 1990;  Melia \& Misra 1993;  Misra \& Melia 1996), thermal and
non-thermal radiation, and the heating of protons via gravitational energy dissipation,
results in a mixture of interacting particles and radiation that produces important 
deviations from a Maxwellian profile for the lepton distribution.  Recently,
Li, Kusunose \& Liang (1996) discussed a stochastic particle acceleration
scheme to account for this high-energy emission, and concluded that the
energized particles consist of a blend of thermal and nonthermal components.
In addition, they found that for certain parameter regimes, acceleration
is more efficient than the cooling processes and must therefore be taken
into account when determining the particle distributions of the emitting
plasma.  Most importantly, they showed that a deviation of the particle
distribution from a Maxwellian at high energies can account for most
of the emission above $1$ MeV via inverse-Compton scattering.  

This alone makes the fully-consistent approach we describe here essential for 
a more complete understanding of the high-energy emission in GBHCs, but there 
are several equally important additional aspects to this problem.  One of 
these is the appearance of a transient $\sim$ MeV bump discovered in Cygnus
X-1 by Ling et al. (1987), which is difficult to explain with the preliminary
study of Li et al. (1996).  Earlier, Melia \& Misra (1993) suggested that
the transient bump might be due to bremsstrahlung-self-Compton processes within
the gravitationally-compressed, inner hot region of the disk.  However, without
a complete analysis of the actual particle distribution, including its 
deviations at high energy, this hypothesis remains to be
tested.  A second important concern is the rapid spectral variability in
these sources, which the {\sl Rossi X-Ray Timing Explorer} is now beginning
to sample with sub-millisecond resolution.  Since the rapid acceleration 
and cooling processes affect the particle energy re-distribution on this
time scale (e.g., Melia \& Misra 1993), it would be highly desirable to have 
a time-dependent scheme to track the particle and spectral evolution in a 
self-consistent manner.   The approach described in this paper will be able
to address each of these issues.

The OSSE instrument on {\sl CGRO} has also detected several Seyfert galaxies
above $60$ keV (e.g., Maisack et al. 1993; Cameron et al. 1995; Madejski 1995),
showing a break or an exponential cut-off at $\sim 50-100$ keV, with an overall
spectrum not unlike those of the GBHCs (e.g., Sunyaev et al. 1991).  Together
with X-ray observations by GINGA, these OSSE data suggest that a single
robust mechanism may be operating in both classes of sources (Fabian 1994).
However, although basic models such as the pure non-thermal pair model
(for a review, see Svensson 1994) predict a steepening of the hard X-ray
spectrum, they do not easily produce a break as sharp as that seen by
OSSE.  In addition, they predict a flattening of the spectrum at $\sim
100-200$ keV, and an annihilation line, neither of which has been
observed so far.  The Seyfert spectra may instead be based on thermal
models, or more realistically, quasi-thermal models (e.g., Haardt \& 
Maraschi 1991, 1993; Ghisellini, Haardt \& Fabian 1993; Titarchuck
\& Mastichiadis 1994;  Zdziarski et al. 1994;  Zdziarski et al. 1995;
Haardt, Maraschi \& Ghisellini 1996; see Svensson 1996 for a recent review).
In the past, purely thermal models for the high-energy spectra of AGNs
have been criticized on the basis of the long thermalization time
needed to attain thermal equilibrium at high temperatures.  The reason 
for this is that the observed X-ray variability time scale is shorter 
than the inferred time required for two-body thermalization, suggesting 
that the plasma cannot be Maxwellian.   Thus, both thermal and non-thermal
models for the X-ray spectra in Seyfert galaxies have significant problems.
In some regions of parameter space, non-thermal and thermal particle
distributions produce very similar spectra, as long as the radiative
process is multi-Compton scattering and the maximum particle energy
is a few MeV (Ghisellini et al. 1993), suggesting that a compromise
model may still work.  However, this similarity in spectra occurs only
for a restricted range of optical depths and source photon compactness,
and the assumed dynamics of the emitting plasma may not be consistent
with the luminosities implied by the observed photon emissivities.  There is clearly
a need to establish the correct (non-Maxwellian) particle distribution 
in these sources for a broad range of system parameters.  In a portion
of the available phase space, the photons are not multiply-scattered 
so the resultant spectrum will in fact reflect the underlying particle 
distribution.  Very importantly, the self-consistently determined 
particle and photon populations must be reconciled with the permitted 
system-dependent considerations, such as the allowed compactness and 
lepton temperature, which turn out to be dependent upon the actual electron
and positron distributions for part of the parameter space.

Our goal in this paper is to begin the process of developing a fully 
self-consistent theory for the interaction of arbitrarily distributed particles 
and radiation to arrive 
at an accurate representation of the high-energy plasma in compact sources.  
We first consider the non-Maxwellian parameter space in \S 2. We then
motivate the use of the Fokker-Planck approach in \S 3, and we derive the general 
Fokker-Planck coefficients for arbitrary particle distributions in \S 4.  In \S 5 
we extend the generality of our scheme by including the electron/positron-proton 
interaction and the effects due to Comptonization (pair creation and annihilation 
are included by means of some standard formulae available in the literature and 
discussed in \S 6).  We test the algorithm in \S 6. Much attention in
that section is paid to the physical consequences of the distribution function
being non-Maxwellian. We end with our conclusions in \S 7. The Appendix
describes in detail the numerical time-dependent scheme that has been used 
to solve the full electron-photon-proton set of equations. The reasons for
neglecting large angle scatterings in the diffusion coefficient are also given
in the Appendix. 

\bigskip
\centerline{\bf 2. The Non-Maxwellian Parameter Space}
\medskip
In this section, we discuss briefly the circumstances under
which one should reasonably expect to see the electron distribution 
deviating substantially from a Maxwellian shape.  Although this 
issue is covered extensively in the literature (e.g., Stepney 1983; 
Guilbert and Stepney 1985; DL89; Baring 1987; 
Ghisellini, Haardt \& Fabian 1993, hereafter GHF; Fabian 1994),
there seems to be no simple expression which would characterize
the parameter space where the non-thermal distribution 
can form for a general source geometry and arbitrary optical depth.
Our goal here is to provide such an expression.

The most likely process to dominate the truncation of the electron 
distribution is inverse Compton cooling on low energy  photons (Fabian 1994).
As is well known (e.g., Rybicki \& Lightman 1979), the rate of
energy transfer in the Thomson scattering regime, from a single electron 
with gamma-factor $\gamma$ to soft photons with energy density $U_{\rm ph}$, is
$$\eqnam{\pcool}
m_e c^2 \,\Bigl ({dE\over dt}\Bigr )^- =
 {4\over 3}\,\sigma_T\,c\,\beta^2\gamma^2\,U_{\rm ph}\;,
\eqno(\new)
$$
where $U_{\rm ph}$ is the soft photon energy density within the source.
For a source with a spherical geometry of radius $R$, one
can approximate this expression as 
$$\eqnam{\plc}
\Bigl ({dE\over dt}\Bigr )^- \simeq
(4/3)\, l \ (\beta\gamma)^2 \, {c(1+\tau_T)\over R}\;,
\eqno(\new)
$$
where $l\equiv L\sigma_T/4\pi R m_e c^3$ is the compactness of the region,
and $L$ is the total source luminosity.
Equation (\plc) does not change for a disk geometry, but the compactness 
$l$ should then be replaced by $l=  L\sigma_T/h m_e c^3$, where $h$ is
the geometrical thickness of the disk, and $L$ is the 
luminosity of a cube $h^3$ (i.e., the standard definition of 
$l$; e.g., Svensson 1996; \S\ 6). 

For the electron-electron energy exchange rate, we will use an
approximation that is well known in the literature and re-derived
below (Eq. 30). This approximation is valid for $E$ ($\gg \Theta_e$) 
equal to the average thermal $\gamma$-factor $\langle\gamma\rangle$:
$$
\eqnam{\elel}
\Bigl ({dE\over dt}\Bigr )_{\rm ee} \simeq
- (3/2)\, {\ln \Lambda\over t_T} \,{1\over \beta \langle
\gamma\rangle}\;. 
\eqno(\new)
$$
Let us now introduce the parameter $\lambda$ which will give the ratio
of the inverse Compton cooling rate to the e-e Coulomb energy transfer rate 
for the average electron:
$$
\eqnam{\comcoul}
\lambda\equiv {|({dE\over dt})^-|\over
|({dE\over dt})_{\rm ee}|}\,\simeq \,
\Bigl ({l \,(1+\tau_T)\over \tau_T\ln \Lambda}\Bigr)\;
 \langle\beta\,\gamma\rangle^3\;,
\eqno(\new)
$$
where $\tau_{\rm T}\equiv n_e\sigma_{\rm T} R$ is the Thomson optical depth.
Considering the relativistic and non-relativistic limits of
$\langle\beta\,\gamma\rangle^3$, and merging these two expressions for
the mildly relativistic domain, one can simplify Equation (\comcoul) to
$$\eqnam{\lsim}
\lambda \simeq\, {l\,(1+\tau_T)\over \tau_T\ln \Lambda}\;
\Bigl ({3\over 2}\,\Theta_e\Bigr)^{3/2}\,\Bigl [
1 + \bigl(6\,\Theta_e\bigr )^{3/2} \Bigr ]\;.
\eqno(\new)
$$

The physical significance of this parameter is such that when
$\lambda$ is of order unity, one should expect a noticeable truncation in the
tail of the Maxwellian distribution. It is {\it not} a quantitative 
parameter that describes exactly how the high energy tail of the 
Maxwellian is truncated, because this is determined not only by  
electron-photon and electron-electron interactions, but also by the 
interaction that heats the electrons.  Note that $\lambda \simgt 1$ does 
not necessarily mean that the photon spectrum will be changed significantly 
due to such a truncation (see the tests in \S\ 6 below).

One can rewrite Equation (\lsim) in terms of the maximum compactness
of the region where e-e Coulomb collisions can still maintain a Maxwellian 
distribution. This gives an expression similar to the one derived by
Fabian (1994), except for the factor $(1+\tau_T)/\tau_T$, which was
approximated by unity for $\tau_T\simgt 1$. Since recently much attention
has been given to optically thin plasmas (e.g., Svensson 1996), 
we will here retain this factor.

Another competitive thermalization process is the so called synchrotron 
boiler (e.g., Ghisellini, Guilbert \& Svensson 1988). It turns out that 
synchrotron photons emitted and re-absorbed by electrons can be a very important
mechanism for the electron thermalization. Although we do not consider this 
thermalization mechanism in this paper, i.e., we assume that magnetic field is
small, it is necessary to asses the importance of this process.
The total synchrotron power emitted by an electron is given by an equation
similar to Equation (\pcool), except that the radiation energy density should be 
replaced by the magnetic field energy density ($U_{\rm mag} = B^2/8\pi$). 
Thus, when the magnetic field energy density is larger than the radiation energy 
density, the electron distribution will be thermalized due to the synchrotron 
self-absorption process or due to Coulomb electron-electron interactions,
if the latter dominate. To see when synchrotron thermalization 
dominates over Coulomb thermalization, let us define a `magnetic' 
thermalization parameter, $\lambda_m$. Following Ghisellini, Guilbert \& 
Svensson (1988), we first define the magnetic compactness parameter $l_b$:
$$\eqnam{\magcom}
l_b \equiv U_{\rm mag} {R \sigma_T\over m_e c^2}\;.
\eqno(\new)$$
The synchrotron self-absorption will dominate as a thermalizing
mechanism if $\lambda_m$, defined as the ratio of the synchrotron emissivity 
to the electron-electron energy exchange rate, is larger than unity:
$$\eqnam{\maglam}
\lambda_m  \simeq\, {l_b\over \tau_T\ln \Lambda}\;
\Bigl ({3\over 2}\,\Theta_e\Bigr)^{3/2}\,\Bigl [
1 + \bigl(6\,\Theta_e\bigr )^{3/2} \Bigr ] \simgt 1\;.
\eqno(\new)
$$
Here we have followed the same steps as those leading to Equation (\lsim).
The physical significance of $\lambda_m$ is that if it larger than unity, then
synchrotron self-absorption thermalization dominates over Coulomb thermalization.
It is interesting to note that recent models for the origin of X-rays in Seyfert
Galaxies invoke short-lived and intense magnetic flares 
on the surface of a cold accretion disk (e.g., Haardt, Ghisellini \&
Maraschi 1994; Nayakshin \& Melia 1997a). In these flares $l_b > l$,
since the conditions
for plasma confinement require that radiation pressure (the dominant pressure
in the case $l\gg 1$) is much smaller than the magnetic stress
(Nayakshin \& Melia 1997b). Therefore, it is very likely that the synchrotron
processes keeps the particles thermal in such flares, unless the heating mechanism
acts in the opposite direction (the heating mechanism is likely to be of a collective
rather than a two-body nature, and it is still unknown; see \S\ 6 below).
In other circumstances, e.g., inside of the accretion disk, the
magnetic field energy density must certainly be below the equipartition value, 
and for the
radiation dominated part of the disk, the `synchrotron thermalization' will be
weaker than inverse Compton cooling. For our purposes here, we
restrict the parameter space of interest to the regions where $\lambda_m$ is 
smaller than $1$, so that Coulomb thermalization dominates.

We should also note that other processes are unlikely to be of a real
importance in determining the shape of the equilibrium electron distribution
(this will be clear from the discussion below).
For example, pair creation and annihilation is a considerably slower process;
we experimented with non-thermal models where pairs are injected with some
power-law distribution (e.g., Lightman \& Zdziarski 1987), and found that
pairs have essentially the same distribution as the electrons everywhere, except
that normalization of the power-law component is different with respect to
the thermalized part of the distribution. Thus, it is fair to assume 
that the electrons and positrons have the same distributions but with different
normalizations. Next, proton heating is also unlikely to change the shape of
the electron distribution because it is strongest at the low energy end.
In that region, the equilibrium distribution when only the
proton-electron interactions remain in the full equation is a Maxwellian
with the proton temperature. But for the low energy end ($E\ll kT$),
the electron distribution function is the same for all temperatures, differing only by
its normalization. Therefore, the low energy end of the electron distribution function
in the presence of proton heating has the Maxwellian shape, irrespective of how
strong the proton heating is. If the electron heating mechanism is different from
proton heating, however, then there is a possibility that it will lead to
deviations from a Maxwellian on the low energy end (where it is the strongest, 
presumably), but we leave this for future work since the exact heating mechanism is
not yet known.
\bigskip
\centerline{\bf 3. The Fokker-Planck Approximation For Coulomb Interactions}
\medskip
As is well known, the Coulomb cross-section diverges for small angle scattering,
where the relative change in the particle's energy approaches a negligible value
(e.g., Chandrasekhar 1942).  It is therefore unwise and 
numerically very challenging 
to treat Coulomb scattering using the otherwise commonly 
employed Boltzmann collision 
integral.  Instead, this feature of the Coulomb force 
lends itself more naturally to 
a Fokker-Planck (FP) approach for solving the kinetic equation in the presence of 
an $r^{-2}$ type of interaction.  The FP procedure is particularly powerful when
the change in particle energy $\Delta E$ is much smaller than its incident energy
$E$.  When this holds, the distribution functions appearing inside the Boltzmann 
collision integral can be decomposed as power series in the small expansion
parameter $\Delta E/E$ ($\ll 1$).  An example of how this works in practice is the 
Kompaneets equation (Kompaneets 1957), which is well-known in astrophysics.

When the particle distribution function is isotropic and 
homogeneous in space, the FP
equation for the distribution function takes the form
$$\eqnam{\fp}
{\partial f(E,t)\over \partial t} = -{\partial\over \partial E}
\Bigl [  a(E,t) f(E,t) \Bigr ]
+ {1\over 2} {\partial^2\over \partial E^2}
\Bigl [  D(E,t) f(E,t) \Bigr ]\;,
\eqno(\new)
$$
where $f(E,t)$ is the distribution function of electrons in energy space (i.e.,
$f(E,t) = \beta \gamma^2 f^{\ast}(E,t)$, $E$ is kinetic 
energy of the electron in units of $m_e c^2$, $E=(\gamma -1)$,
$\beta\equiv (1-1/\gamma^2)^{1/2}$ and $f(E,t)dE
\equiv f^{\ast}(\vec p,t) d^3\vec p$ gives the volume density  of 
electrons in the phase space element $d^3\vec p\,$, and $\vec p$ is the electron
momentum in units $m_e c^2$),
$a(E,t)$ is the average energy exchange rate and $D(E,t)$ is the energy dispersion 
rate of the test particle. Both $a(E,t)$ and $D(E,t)$ 
are functions of time because the 
distribution function over which they are convolved depends on time. Since $E$ is
dimensionless, we will use $E$ and $(\gamma-1)$ interchangeably throughout the paper, 
and $f(\gamma)$ is the same distribution as $f(E)$.

In their application of this equation to the study of 
electron thermalization in
a background relativistic Maxwell-Boltzmann (MB) plasma, 
Dermer and Liang (1989)
derived the FP coefficients for a small population of 
particles scattering off a 
relativistic MB distribution.  In this limit where 
the test particle number density 
is much lower than that of the background MB plasma, 
Equation (\fp) is linear in the
distribution function $f(E,t)$.  However, this approach 
fails when the real electron
distribution differs substantially from the perfect (background) MB profile.
The original motivation for taking 
this simplified approach was to produce one-dimensional
integrals for the FP coefficients.  However, as we shall see in \S 4 below, the
FP coefficients derived there for an arbitrary particle distribution 
are themselves simply one-dimensional
integrals and there is therefore no computational 
advantage to be had in restricting the
problem to the quasi-thermalized initial condition if the temperature is
allowed to evolve.  Of course, our generalized approach
is practical only if we can find a numerical scheme 
that makes the problem tractable.  The nature of the computational scheme 
that solves this problem is discussed in detail in the Appendix.
\vfill\eject\null
\centerline{\bf 4. Coulomb Electron-Electron Fokker-Planck Coefficients}
\centerline{\bf For Arbitrary Particle Distributions}
\medskip
\centerline{\sl 4.1. The Single Particle Energy Exchange $a(\gamma,\gamma_1)$
and Diffusion $D(\gamma,\gamma_1)$ Coefficients}
\medskip

We begin by deriving the Fokker-Planck coefficients for e-e Coulomb
interactions for
a mono-energetic particle distribution. The FP
coefficients for an arbitrary particle profile can be calculated
from these by convolving them over $f(E,t)$.
We consider the FP energy exchange and diffusion coefficients for
a test particle with dimensionless energy $E\equiv (\gamma - 1)$ (all 
electron energies are in units of $m_e c^2$ in this paper) Coulomb
scattering off a mono-energetic electron (positron) distribution
with energy $E_1\equiv (\gamma_1-1)$. Our notation and some
of our starting equations are based on earlier work by
Dermer (1985) and Dermer \& Liang (1989) (DL89 hereafter). 

The distribution function is normalized by the condition 
$$\eqnam{\norma}
1=\int\limits_{1}^{\infty} d\gamma\, \beta\gamma^2 f^{\ast}(\gamma)
\equiv\int\limits_{1}^{\infty} d\gamma\,  f(\gamma) \;.
\eqno(\new)
$$
For two such functions $f^{\ast}_1(\gamma_1)$ and $f^{\ast}_2(\gamma_2)$,
the relativistically correct reaction rate is 
$$\eqnam{\rate}
r={ c n_1 n_2\over 2(1+\delta_{12})}\int\limits_{1}^{\infty}
\,d\gamma_r\, (\gamma_r^2-1)\,\integral\, d\gamma_c\, 
(\gamma_c^2-1)^{1/2}\int\limits_{-1}^{1}\,du\, f^{\ast}_1
(\gamma_1)\,f^{\ast}(\gamma_2) \sigma(\gamma_r,\gamma_c)\;,
\eqno(\new)
$$
where $n_1$ and $n_2$ are the corresponding particle number
densities, $\gamma_r$ is the invariant relative Lorentz factor and
$\gamma_c$ is the Lorentz factor associated  with velocity of
the center of momentum (CM) frame relative to the lab frame
(Dermer 1984, 1985).  Here, $u\equiv\cos\mu$, and the variable 
$\mu$ is defined as the angle between the velocity of the CM 
frame and the particle 1 velocity in that frame. The Kronecker 
delta $\delta_{12}$ prevents double counting.

To arrive at the relativistically correct reaction rate for two 
mono-energetic distributions, we take the particle functions 
$f^{\ast}_i(\gamma_1)$, $i=1,2$, to be 
$$
\eqnam{\funct}
f^{\ast}_i(\gamma_i)={1\over \beta_i \gamma_i^2}\,\delta(\gamma_i'-\gamma_i)\;,
\eqno(\new)
$$
where $\beta_i\equiv (1-1/\gamma_i^2)^{1/2}$,
$\gamma_i'$ are  the laboratory frame Lorentz factors of
the colliding particles expressed in terms of the relative
Lorentz factor $\gamma_r$, and the CM Lorentz factor $\gamma_c$: 
$$\eqnam{\center}
m_{1(2)}\,\gamma_{1(2)}'= {\gamma_c\,\over S^{1/2}}
\Bigl [ (m_{2(1)}\gamma_r+m_{1(2)})
\pm \beta_c m_{2(1)}\beta_r \gamma_r\,u\Bigr ]\;,
\eqno(\new)
$$
where $S\equiv{m_1^2+m_2^2+2m_1m_2\,\gamma_r}$ and $+$ is
for $i=1$, while $-$ sign is for $i=2$.

In deriving the energy exchange coefficient, we use inside the
integral of Equation (\rate) the energy exchange rate 
$\langle\sigma(\gamma_r,\gamma_c)\Delta E\rangle$ 
for Coulomb interactions 
averaged over all {\it small} (see below)
scattering angles in the center of momentum frame
given explicitly by
$$\eqnam{\enrate}
\langle\sigma(\gamma_r,\gamma_c)\Delta E \rangle =
\int d^3  \vec {p^{*}}\; {d\sigma\over d \vec {p^{*}}^3}
\;\Delta E(\vec {p^{*}})\;,
\eqno(\new)
$$
where $\Delta E$ is the energy exchange per scattering
(in the laboratory frame) and $\vec{p^{*}}$
is the particle momentum in the center of momentum frame. 

Before we proceed with the derivation, we note that it only makes 
sense to talk about FP coefficients for the Coulomb 
interactions for small scattering angles. 
Even though such interactions dominate in the Coulomb scattering
cross section, one must explicitly exclude the large
scattering angles in the integrals for the
FP coefficients. We will show later that inclusion of 
these angles in the FP coefficients  (as done by DL89, for example)
leads to erroneous results.
Instead, the large angle scatterings must be included by
means of the exact Boltzmann equation, that is the original integral
equation from which the FP equation is derived. Following the standard
practice, we neglect the large scattering
angles in  the Coulomb interactions in this paper.

Using the notation of Dermer (1984, 1985), the change in electron energy is 
$$\eqnam{\echange}
\Delta E = \gamma_c\, \beta_c\,p^{*} \Bigl [
(\cos \Psi^* - 1)\cos \mu- \sin\Psi^*\cos\phi^*\sin\mu
\Bigr ],
\eqno(\new)
$$
where $\Psi^{*}$ is the
scattering angle in the center of momentum system, and $\phi^*$ 
is the polar angle in the C-system of coordinates (i.e., the CM frame in
which the $z$-axis is directed along the unscattered particle 1 momentum;
see Dermer 1985, Fig.1).  Because the CM frame 
Lorentz $\gamma$-factor $\gamma^{*}$ of the colliding
electrons is $\gamma^{*}=\sqrt{(\gamma_r + 1)/2}$
for particles of equal mass, and $p^*=\beta^*\,\gamma^*$, 
the integration in Equation (\enrate)
is only carried out over angles $\phi^*$ from 0 to $2\pi$ and $\Psi^*$ from
$\Psi^*_{\rm min}$ to some $\Psi^*_{\rm max}$ (see below).

Since we limit ourselves to the small scattering angles, $\Psi^*
\ll 1$, we can leave only the leading terms in the Coulomb
scattering cross section, which is then equal to 
$$\eqnam{\sech}
{d\sigma\over d \cos\Psi^{*}}=
{2\pi {r_e}^2 (2 {\gamma^{*}}^2-1)^2
\over {\gamma^{*}}^2({\gamma^{*}}^2-1)^2 }
\;{1\over (1-\cos^2\Psi^{*})^2}
\eqno(\new)
$$
(see, e.g., Jauch \& Rohrlich 1980), where $r_e$ is the classical electron radius. 

The exact value of $\Psi^*_{\rm max}$
is usually considered to be unimportant and therefore often chosen 
as $\pi/2$ for simplicity (e.g., Landau and Lifshitz 1981; Dermer 1985).
Because the second term inside the brackets in Equation (\echange) gives
zero after averaging over $\phi^*$, the energy exchange 
rate in the CM frame is then
$$\eqnam{\enxch}
\langle\sigma(\gamma_r,\gamma_c)\Delta E\rangle =
\int_{\cos\Psi^*_{\rm min}}^{\cos\Psi^*_{\rm max}}
d\cos\Psi^* {d\sigma\over d \cos\Psi^{*}}
\gamma_c\beta_c p^* 
\Bigl [
(\cos \Psi^* - 1)\cos \mu \Bigr ]{\rm .}
\eqno(\new)
$$
Taking the integral and expressing $p^*$ in terms of
$\gamma_c$ and $\gamma_r$, we obtain
$$\eqnam{\aa}
\langle\sigma(\gamma_r,\gamma_c)\Delta E\rangle = 8\,
\pi {r_e}^2\, \ln\Lambda\;{ {\gamma_r}^2\gamma_c\beta_c
\over \sqrt{2(\gamma_r+1) (\gamma_r-1)^3}}\; u\;,
\eqno(\new)
$$
where $\ln\Lambda\equiv\ln\Bigl [(1-\cos\Psi^*_{\rm max})
/ (1 - \cos\Psi^*_{\rm min})\Bigr ]^{1/2}$ 
is the Coulomb logarithm.
We set $\ln \Lambda = 20$ throughout the paper.

From Equation (\rate), we now infer the general expression for the 
energy exchange coefficient (note that this expression is now for
a single electron, so $n_2$ no longer enters the expression; also
the factor with the Kronecker $\delta_{12}$ was discarded since we
now are interested in the energy exchange rate for a given particle rather
than the total interaction rate):
$$\eqnam{\ratea}
a(\gamma,\gamma_1)=
{ c n_1 \over 2\beta\gamma^2\beta_1\gamma_1^2}\int\limits_{1}^{\infty}
\,d\gamma_r (\gamma_r^2-1)\,\integral\, d\gamma_c (\gamma_c^2-1)^{1/2}
\int\limits_{-1}^{1}\,du\, 
\delta\{\gamma-\gamma_2'\}\, 
\delta\{\gamma_1-\gamma_1'\}\, 
\langle\sigma(\gamma_r,\gamma_c)\Delta E\rangle\;.
\eqno(\new)
$$
Performing the third integration first, one finds
$$\eqnam{\aab}
\int\limits_{-1}^{1} d u\, 
\delta\{\gamma-\gamma_2'\}
\langle\sigma(\gamma_r,\gamma_c)\Delta E\rangle = -
{8\pi r_e^2 \kyl\over m_e^2} {S \gamma_r^2\sqrt{(\gamma_r-1)/2}
\over(\gamma_r^2-1)^2 (\gamma_r-1) \gamma_c
\beta_c} \,\left(\gamma-\gamma_c{\gamma_r+1 \over (S^{1/2}/m_e)} \right )\;, 
\eqno(\new)
$$
subject to the condition that $|\cos\mu| \equiv |u| \leq 1$, which
leads to the constraint
$$\eqnam{\aac}
{(S^{1/2}/m_e) \over \gamma_r \beta_r \gamma_c \beta_c}
\left |\gamma-\gamma_c {\gamma_r + 1\over (S^{1/2}/m_e)}
\right |\leq 1\;.
\eqno(\new)
$$
According to Equation (\center), $\gamma_c$ is fixed by the expression 
$\gamma_c=(\gamma+\gamma_1)/{(S^{1/2}/m_e)}$ when $m_1 = m_2 
= m_e$.  Equation (\aac) then leads to 
a restriction on the range of physically accessible values of
$\gamma_r$ in the integral of Equation (\ratea), which we write as
$\gamma^- \leq\gamma_r\leq \gamma^+$, where $\gamma^{\pm}=\gamma_1
\gamma(1\pm \beta_1 \beta)$.

The next step is to compute the integral over $\gamma_c$:
$$\eqnam{\aad}
\integral d\gamma_c \gamma_c\beta_c
\int\limits_{-1}^{1} d u\,  
\delta\{\gamma-\gamma_2'\}\, 
\delta\{\gamma_1-\gamma_1'\}\, 
\langle\sigma(\gamma_r,\gamma_c)\Delta E\rangle = -
{4\pi r_e^2 \kyl \over
(\gamma_r^2-1)^{3/2}(\gamma_r-1)}\,(\gamma-\gamma_1)\;.
\eqno(\new)
$$
The final integral leads to the energy exchange coefficient, which we write as
$$\eqnam{\ab}
a(\gamma,\gamma_1) = -{2\,\pi {r_e}^2c n_1 
\kyl\over \beta_1{\gamma_1}^2\beta\gamma^2 }\,
\bigl (\gamma - \gamma_1 \bigr )\,\chi(\gamma,\gamma_1)\;,
\eqno(\new)
$$
where $\chi(\gamma,\gamma_1)$ is the integral function defined by
$$\eqnam{\ac}
\chi(\gamma,\gamma_1) = \int\limits_{\gamma^{-}}^{\gamma^{+}}
dx\;\, {x^2\over \sqrt{(x+1)(x-1)^3\,}}\;.
\eqno(\new)
$$
This integration can be carried out analytically, giving
$$\eqnam{\ad}
\chi(\gamma,\gamma_1) = \left [-\sqrt{{x+1\over x-1}}
+ 2 \sinh^{-1}{\Bigl (\sqrt{{x-1\over 2}}\Bigr )} 
+ \sqrt{x^2-1}
\right ]^{\gamma^+}_{\gamma^-}.
\eqno(\new)
$$
Note that since $\gamma^{\pm}$ is symmetric in 
$\gamma$ and $\gamma_1$, the energy 
exchange rate given by Eq.(\ab) is antisymmetric
 with respect to the interchange 
$\gamma\leftrightarrow\gamma_1$.

For the diffusion coefficient, we proceed in a similar fashion.
Now, however, it is imperative (see Appendix A.4)
that we  choose $\Psi^*_{\rm max}\ll 1$. We square Equation
(\echange) and consider the term with $(1-\cos\Psi^*)^2$.  If
one takes $\Psi^*_{\rm max}=\pi /2$, angles $\Psi^*\sim 1$ will
dominate the contribution of this term to the integral over $\Psi^*$. 
But this is the  parameter space where the FP equation becomes
invalid, thus we cannot simply suggest $\Psi^*_{\rm max}=\pi /2$
and must use $\Psi^*_{\rm max}\ll 1$. In that case, contribution of the
first term in Equation (\echange) is negligible compared to the
second term in the brackets. Accordingly, 
$$
\eqnam{\ae}
\langle\sigma(\gamma_r,\gamma_c)(\Delta E)^2\rangle = 
{4\pi {r_e}^2 {\gamma_r}^2({\gamma_c}^2-1)
\over {\gamma_r}^2-1}\,\kyl\,\bigl (1- u^2\bigr )\;,
\eqno(\new)
$$
and after carrying out the necessary integrations, we arrive at
$$\eqnam{\af}
D(\gamma,\gamma_1) = {4\,\pi {r_e}^2c n_1 
\kyl\over \beta_1{\gamma_1}^2\beta\gamma^2 }\,
\Bigl [{1\over 2}\,(\gamma-\gamma_1)^2\,\chi(\gamma,\gamma_1)
+\zeta(\gamma,\gamma_1)
\Bigr ]\;.
\eqno(\new)
$$
The function $\zeta(\gamma,\gamma_1)$ is defined as
$$\eqnam{\ag}
\zeta(\gamma,\gamma_1) = \int\limits_{\gamma^{-}}^{\gamma^{+}}
dx\, {x^2\over \sqrt{x^2-1}}
\Bigl [{(\gamma+\gamma_1)^2 \over 2 (1+x)} - 1 \Bigr ]\;.
\eqno(\new)
$$
The integrand is a relatively well behaved function, and the integral 
is easy to evaluate numerically as well as analytically.

Figures 1 and 2 show the FP energy exchange $a(\gamma,\gamma_1)$
and dispersion $D(\gamma,\gamma_1)$ coefficients, respectively,  as functions
of $\gamma$ for three different values of $\gamma_1$.  In these
figures, $t_C$ is the Thomson time divided by the Coulomb logarithm ($t_C
\equiv 1/n_e c \sigma_{\rm T} \ln \Lambda$).
It is evident that $a(\gamma,\gamma_1)$ is a discontinuous function of
$\gamma$;  at $\gamma=\gamma_1$ it changes sign from a
non-zero positive value to a non-zero negative one.
This non-physical behavior is caused by the
divergence of the original Coulomb scattering cross section 
for $|\vec v_1 - \vec v_2| \ll v_1$. This
discontinuity in $a(\gamma,\gamma_1)$
 for $|\vec v_1 - \vec v_2| \ll v_1$ is not a problem
physically, because the parameter space for such interactions is very small
and will contribute negligible to the electron-electron 
Coulomb interactions. It  is also never a problem numerically.
The convolution of $a(\gamma,\gamma_1)$ over any real 
physical distribution (i.e., a reasonably smooth one)
smoothes out this non-physical behavior and practically eliminates
the problem (one can show that electrons with energies $|\gamma_1 -
\gamma| = \Delta E$ have $a(\gamma,\gamma_1)\propto (\Delta E)^2$ for
any continuous $f(\gamma)$).

A well known fact is seen from Figure 1: 
particles with lower energy both gain and lose
energy (depending on how energetic their interacting partners are)
faster than the more energetic ones. This happens because of the
dependence of the Coulomb scattering cross section on energy 
of the colliding particles in the CM frame (Eq. 14).

It is possible to derive analytical limits for the FP coefficients.
For $\beta_1 \ll 1$ one gets from Equation (\ab) the limiting form
for $a(\gamma,\gamma_1)$:
$$\eqnam{\alimit}
a(\gamma,\gamma_1) \; = \;\cases{  (3/2)\; {1/ t_C \beta_1},
&\hbox{for} $\; \gamma< \gamma_1$\cr
\noalign{\vskip 0.3cm}
 - (3/2) \; {1/ t_C \beta},
&\hbox{for}$\; \gamma> \gamma_1$\cr}\;.
\eqno(\new)
$$

\noindent In the limit $\gamma\gg\gamma_1\gg 1$ it is also possible to have 
an analytic approximation for $a(\gamma,\gamma_1)$:
$$\eqnam{\blimit}
a(\gamma,\gamma_1)\simeq\; -(3/2)\; {1\over t_C} \;
\Bigl( {1\over \gamma_1} -  {1\over \gamma}\Bigr )
\; \simeq\; -(3/2)\; {1\over t_C}\;{1\over \gamma_1}\;.
\eqno(\new)
$$
We can merge these two expressions in the case of 
$E \equiv \gamma - 1 \gg E_1 \equiv  \gamma_1 - 1 $
to give 
$$
\eqnam{\climit}
a(\gamma,\gamma_1) \; = \; -\,(3/2) \; {1\over t_C} 
\; {1 \over \beta \gamma_1}
\eqno(\new)
$$

\medskip
\centerline{\sl 4.2. FP Coefficients For Arbitrary Particle Distributions: Examples}
\medskip

Now that we have the mono-energetic coefficients in hand, finding
those for any given arbitrary particle distribution $f(\gamma_1)$
is simply a matter of convolving these with the particle profile:
$$\eqnam{\any}
a(\gamma) = \int d\gamma_1 f(\gamma_1) a(\gamma,\gamma_1)\;,
\eqno(\new)
$$
and similarly for $D(\gamma)$.

To illustrate how this works in practice, we show in Figures 3 and 4 
the e-e Coulomb energy exchange and diffusion coefficients for three 
different distributions: (1) a perfect Maxwellian at temperature 
$\Theta_e = 1$ , (2) a Gaussian with the same average energy, and (3)
a power-law $f(\gamma)=\beta\gamma^{-p}$ (with the maximum 
$\gamma_{\rm max} - 1 =100$, $\gamma_{\rm min}-1 = 0.007$  and $p = 2.48$).
  The power-law and the Gaussian are chosen as two
opposite representative cases, the former broader than and the
latter narrower than the Maxwellian.  All three distributions 
are normalized to unity and are plotted in Figure 5.

We note that the zeros of the energy exchange coefficient occur not 
at $\gamma=\langle\gamma_1\rangle$, where $\langle\gamma_1\rangle$ 
is the average Lorentz
factor of the distribution, but rather at the maximum of the number 
distribution function.  It is evident also that the quantity 
$\int d \gamma |a(\gamma)|$ is larger for the more diffuse distributions.
The diffusion coefficient behaves in the opposite way. The larger diffusion coefficient
corresponds to the  sharper  distribution function.
Summarizing, we conclude that if the distribution is broader than a
Maxwellian, it will be brought to the thermal distribution by the
energy exchange coefficient, whereas if it is narrower,
it will reach the thermal distribution as a result of the
diffusion coefficient.

Dermer and Liang (1989) derived the energy exchange and the diffusion
coefficients for the Coulomb interactions of a test electron with
a background Maxwell-Boltzmann electron plasma.
For the Maxwellian distribution, our results for the energy exchange 
coefficient are in complete agreement (to $1/\kyl$ precision)  
with those of DL89. 
For the diffusion
coefficient, however,  our expression disagrees with theirs at both the
high and low energy end.
Comparing the two expressions (Boettcher, 1996, private 
communication) one finds that they 
agree with each other well in the region $E\sim \Theta_e$ (up to the same precision
of $1/\ln\Lambda$), but seriously disagree for $E\ll\Theta_e$ and $E\gg\Theta_e$.
We find that $D(\gamma,\Theta_e) \simeq \hbox{const}$ ($\approx 
2 \Theta_e |a(\infty,\Theta_e)|$) for
$E\gg 3 \Theta_e$, while DL89 found that
$D(\gamma,\Theta_e) \simeq {\hbox{const}}^{\prime} \times E$ in that limit. 
We also find $D(\gamma, \Theta_e)\propto \beta^{1/2}$ on the low energy end, 
whereas DL89 found $D(\gamma, \Theta_e)\propto {\rm const}$.
The difference results from the fact that DL89 used $\Psi^*_{\rm max}=\pi /2$,
and also left the terms leading to large scattering angles in the
Coulomb cross section (see Appendix A.4)
Taking into account the fact that $a(\gamma,\Theta_e) \simeq -
\hbox{\rm const}$ for $E\gg 3 \Theta_e$, we see from Equation (\fp) that the 
equilibrium Maxwellian distribution can be reached only for $D(\gamma,\Theta_e) 
\simeq \hbox{\rm const}$ at the high energy end. If instead,
 $D(\gamma,\Theta_e) \simeq {\hbox{\rm const}}^{\prime} \times E$
were the correct behavior, then one would find that the equilibrium
$f(\gamma)$ is a power-law in the high energy limit. Since the 
original Boltzmann collision integral for Coulomb interactions 
leads to a Maxwellian electron
distribution, this property must be preserved by the FP equation.

\medskip
\centerline{\sl 4.3 The Thermalization Process and Time Scales}
\medskip
An interesting question to consider is the following: 
Is self-consistent thermalization faster or slower than the
thermalization of a test particle distribution interacting with
a (fixed) background Maxwell-Boltzmann plasma?  The answer is 
not obvious because both of the FP coefficients play a role in 
establishing the Maxwellian distribution, and as we saw in the
previous section, if a distribution's energy exchange coefficient 
is larger than that of the Maxwellian, the diffusion coefficient 
is smaller, and vice versa.
  
To answer this question, at least illustratively, we consider the
thermalization of a Gaussian electron distribution in two special
cases: (a) when the electrons interact with each other (i.e., solving 
the fully non-linear problem), and (b) when they thermalize on a 
background MB gas with the same average energy as that of the Gaussian 
distribution.  In order to characterize the interaction quantitatively,
we define the thermalization time scale $t_r$ to be the 
time it takes the particle population to reach a deviation of $5\%$ or
less from a perfect Maxwellian, where the deviation $\varepsilon$ is 
defined to be 
$$\eqnam{\deven}
\varepsilon\, \equiv\, {\integral\,d\gamma\,(\gamma-1)\,
|f(\gamma,t)- f_M(\gamma)|\over \integral
\,d\gamma\,(\gamma-1)\,f_M(\gamma)}\;.
\eqno(\new)
$$
In this expression, $f(\gamma,t)$ and  $f_M(\gamma)$ are, respectively,
the actual time-dependent distribution function and the perfect 
Maxwellian with the same average energy and number of particles.

Our tests indicate that to within $\sim 5\%$ accuracy, the
time scale for the test particle thermalization coincides with that
for the self-interacting case.  The explanation for this appears to
be that most of the elapsed time occurs during the period when the
distribution is close to Maxwellian, when in fact the FP
coefficients for the time-dependent distribution and those for the Maxwellian
are very similar.  As such, our results are in very good agreement
with the findings of DL89:
$$\eqnam{\stepney}
t_r\,\, = \, 4 t_T \Theta_e^{3/2}(\pi^{1/2}- 1.2 \Theta_e^{1/4}
+2 \Theta_e^{1/2})/\kyl\;,
\eqno(\new)
$$
where $t_T= (n_1 c \sigma_T)^{-1}$ is the Thomson time.
 At the same time
 we find that the thermalization time depends 
on the shape of the initial distribution. However, if one
keeps the shape of the initial distribution fixed (e.g., a power-law)
 and observes the temperature dependence of the 
thermalization time,
it is practically identical to Equation (\stepney).

Figures 6 and 7  show the
time evolution of the initial distribution functions, 
a power-law and a Gaussian, respectively, 
as they approach the equilibrium Maxwellian. The equilibrium 
temperature is the same for both distributions and equals 0.3 $m_e c^2$.
Figure 8 shows the time dependence of the deviation $\varepsilon$
for both cases. It is notable that thermalization of the 
Gaussian happens much faster than that of the power-law. (When
comparing the figures for these two cases, one should 
bear in mind the fact that
different parts of the distribution function contribute to
the integral in Eq. (\deven) with different weights, namely, $E^2$. Thus, the
high energy end of the electron distribution contributes much more to
the deviation $\varepsilon$ than the low energy end.)
To explain this fact, we should consider the Coulomb interaction time scale
for individual particles. Let us define the particle energy time scale $t_e$
as 
$$
\eqnam{\tai}
{1\over t_e(\gamma)} = {1\over E} \; \Bigl [\; |a(\gamma)| + {D(\gamma)\over 2E}\Bigr ]\;,
\eqno(\new)
$$
where $a(\gamma)$ and $D(\gamma)$ are  the Maxwellian FP 
coefficients for the equilibrium electron temperature. 
The physical sense of $t_e$ is that it is the
time it takes to change the particle energy considerably (say, by a factor of 2). 
To obtain the thermalization time scale for a distribution 
as a whole, one should average $1/t_e(\gamma)$ over the
distribution and take the inverse of this value.
Applying this to the two thermalization processes considered (Figs. 6 \& 7),
the power-law has substantially more particles
at the higher energy end than the Gaussian, and it therefore
has a longer thermalization time, in accordance with Figure 8.

Figure $8$ also shows that the equilibrium distribution
function does not exactly coincide with the Maxwellian 
since $\varepsilon$ does not go
to zero when $t\rightarrow\infty$, but rather stays
constant at $\sim 10^{-2}$.
This occurs because the differencing scheme (see Eq. 2),
modifies the true FP coefficients in order to preserve the energy and 
number of particles (see Appendix). 
One can introduce more elaborate energy-dependent corrections to the FP
coefficients, and it will likely reduce the deviation $\varepsilon$ by 
an order of magnitude or more. 
However, since a precision of about 0.01 is satisfactory for our goals,
we will leave these improvements to the future.
\bigskip
\centerline{\bf 5. Collision Integrals For Electron-Proton 
and Electron-Photon Interactions}
\medskip
\centerline{\sl 5.1. Fokker-Planck Coefficients For Electron-Proton 
Interactions}
\medskip
The fact that electrons and protons are subject to the same
Coulomb force as electrons with electrons and electrons with
positrons (other than for the cross section) allows us to use
the same Fokker-Planck approach as described above to account for
these interactions.  The FP coefficients for electron-proton (e-p 
hereafter)
scatterings will differ from those of lepton-lepton interactions
for the additional reason that the particle masses are no longer
the same.  In this section, we shall derive the e-p
energy exchange $a_p(\gamma,E_p)$ and diffusion $D_p(\gamma,E_p)$ 
coefficients along the lines established above ($E_p$ is
the proton kinetic energy in units of $m_p c^2$).  Unlike earlier
work in which the FP coefficients were derived for a thermal
proton distribution (DL89), we here solve the 
problem for an arbitrary proton profile.

We follow the same strategy as above and neglect terms of order 
lower than $\kyl$ in the final expression.  Assuming infinitely 
massive scatterers (the protons), the e-p
differential cross section in the center of momentum frame is
$$\eqnam{\kylprot}
{d \sigma^*\over d\cos \Psi^*}=
{\pi r_e^2\over 2} \,\,\,{1\over
\beta_r^4 \gamma_r^2 \sin^4 (\Psi^*/2)}
\eqno(\new)
$$
(see Jauch \& Rohrlich 1980; we neglect 
the factor $(\beta_r\sin\Psi^*)^2$ compared to
unity in the numerator, since we again only include $\Psi^*\ll 1$).
The validity of this treatment
is limited to $\gamma_r\ll m_p/m_e$, where $m_p$ is the proton
mass.  The mono-energetic e-p energy exchange coefficient
is found to be
$$\eqnam{\aproton}
a_p(\gamma,\gamma_p) =
{2 \pi r_e^2 c n_p \kyl m_e
\over \beta \gamma^2\beta_p \gamma_p^2}
\int\limits_{\gamma^-}^{\gamma^+}
{d \gamma_r \,\gamma_r^2 S^{-1}_p \over  (\gamma_r^2-1)^{3/2} }
\,\, \Bigl [ (\gamma_r-1) (m_p \gamma_p -m_e \gamma)
\, + (m_p+m_e) (\gamma_p - \gamma) \Bigr]\;,
\eqno(\new)
$$
where $\gamma_p$ and $\gamma$ are the proton and electron
Lorentz factors, respectively, $\gamma^{\pm}\equiv 
\gamma \gamma_p (1\pm \beta \beta_p)$, and $S_p\equiv
m_p^2 + m_e^2 +2m_p m_e \gamma_r$.   We note
that due to the difference in electron and proton masses,
$a_p(\gamma,\gamma_p)\not= 0$ even when $\gamma =\gamma_p$.
The energy exchange  coefficient $a_p(\gamma,\gamma_p)$ is shown in Figure 9 
as a function of electron energy, for three different values
of the proton energy. Note that there is still
a discontinuity at $\gamma = \gamma_p$, however.

The corresponding e-p diffusion coefficient is
given by
$$
D_p(\gamma,\gamma_p) =
{2 \pi r_e^2 c n_p \kyl m_p m_e
\over \beta \gamma^2\beta_p \gamma_p^2}
\int\limits_{\gamma^-}^{\gamma^+}
{d \gamma_r \, \gamma_r^2\, S_p^{-1} \over  (\gamma_r^2-1)^{1/2} }\,\times
$$
$$\eqnam{\dproton}
\,\, \left [{ (m_p \gamma_p +m_e \gamma)^2\over
S_p } \,-1
\, - {  [ (\gamma_r-1) (m_p \gamma_p -m_e \gamma)
+ (m_p+m_e)(\gamma_p - \gamma)]^2 \over S_p (\gamma_r^2-1)} \right ]\;.
\eqno(\new)
$$

To get the e-p FP coefficients for an arbitrary 
proton distribution $f_p(\gamma_p)$, one integrates Equations 
(\aproton) and (\dproton) over $f_p$:
$$\eqnam{\arbit}
a_p(\gamma) = \integral d \gamma_p\,
 a_p(\gamma,\gamma_p) \, f_p(\gamma_p)\;.
\eqno(\new)
$$
We plot the e-p diffusion
coefficient for Maxwellian protons with a temperature $k T_p = $
10 MeV and 100 MeV in Figure 10. For comparison, 
we also show the electron-electron diffusion coefficient 
for a Maxwellian distribution with $k T_e = $ 1 $m_e c^2$. 
A graph of the the e-p energy exchange coefficient can be found in
DL89.
Note that the e-p FP coefficients differ from
those of the electron-electron interaction roughly by a factor
$(m_e/m_p) T_p/T_e$. 

\bigskip
\centerline{\sl 5.2. Electron-Photon Interactions}
\medskip

In this section, we present the electron-photon
collision integral corresponding to Compton scattering.
This part of the kinetic equation has already been dealt with
extensively in the literature (see references below in this section),
 especially for isotropic 
photon and particle distributions.  With the assumption of
isotropy in both the electron and photon spectra,
the electron-photon part of the collision integral is
$$\eqnam{\compt}
\Bigl ({\partial f(\gamma,t)\over \partial t}\Bigr )_{e\gamma}=
-f(\gamma)\int d\omega\, N(\omega) R(\omega,\gamma)
+ \int \int d\omega'\, d\gamma'\, R(\omega',\gamma')
P(\gamma;\gamma',\omega') N(\omega') f(\gamma')\;,
\eqno(\new)
$$
where $R(\omega,\gamma)$ is the scattering rate between
photons of energy $\omega$ and electrons of energy $\gamma$;
$P(\gamma;\gamma',\omega')$ is the probability of scattering
an electron with initial energy $\gamma'$ to a
final energy $\gamma$ in a collision with a photon of energy
$\omega'$, and $N(\omega)\,d\omega$ is the number of
photons per unit volume with energy between $\omega$ and
$\omega+ d\omega$.  For the angle-averaged scattering rate 
$R(\omega,\gamma)$, we use the expression given in Equation
(2.3) in Coppi \& Blandford (1990).

The probability $P(\gamma;\gamma',\omega')$
was first derived by Jones (1968) and later corrected by
Coppi \& Blandford (1990).
We carried out the last integral in the expression given
by Coppi \& Blandford (1990) to get $P(\gamma;\gamma',\omega')$ 
integrated over all angles. We do not give the rather
lengthy expression since more complete work exists in the
literature (e.g., Nagirner and Poutanen 1994 and references therein).
 We checked the validity of our
expression by comparing it with the numerical integrations of 
Coppi \& Blandford (1990) and by integrating $P(\gamma;\gamma',
\omega')$ over all the accessible values of $\gamma$,
which should give unity.

A problem that may arise when Equation (\compt) is integrated
numerically is that the electron energy bin size is finite, which
can lead to a situation where scatterings in the low energy
regime cannot correctly transfer the particles in energy space.
Consider a photon with energy $\omega$ (in units of $m_e c^2$)
scattering in the Thomson limit, so that $\omega\gamma^2\,\ll 1$.
The photon emerges with energy $\omega' \sim \gamma^2 \omega$ after
one scattering off an electron with a Lorentz factor $\gamma$. 
If the change $\Delta \omega= \omega' - \omega$ in the photon energy
is smaller than the electron energy bin size, such a scattering will 
`fall through' the energy space mesh. When the density of low energy
photons is large, the error introduced by this effect can be large.

Because of limited computer resources, it is often impractical to simply
reduce the energy bin size.  Instead, the photon energy range used
in Equation (\compt) can be divided into 2 sections. 
Let us define a break energy $\omega_b(\gamma)$ in the photon spectrum
by the expression $\omega_b(\gamma) = \min\{{1\over 2}\, (\gamma-1),\; 3/(4\gamma)\}$.
For low energy photons, 
that is for $\omega< \omega_b$, we introduce the alternative cooling rate
$$
\eqnam{\coolin}
\dot \gamma_C(\gamma) \equiv
\int\limits_{0}^{ \omega_b(\gamma)}
d\omega \, N(\omega) R(\omega,\gamma)\;(\langle\omega_s(\gamma)\rangle-\omega)\;,
\eqno(\new)
$$
where $\langle\omega_s(\gamma)\rangle$ is the mean scattered energy 
of the photon $\omega$
interacting with an electron with Lorentz factor $\gamma$. 
This cooling rate is in fact the electron-photon energy exchange rate
and is analogous to the quantities $a(\gamma,\gamma_1)$ and
$a_p(\gamma,\gamma_p)$. 

One must also take into account the electron diffusion
associated with Compton scatterings, that is the full diffusion
coefficient must include a contribution from electron-photon
scatterings as well. 
We define the electron-photon diffusion coefficient as
$$
\eqnam{\photdif}
D_C(\gamma)  \equiv
\int\limits_{0}^{ \omega_b(\gamma)}
d\omega \, N(\omega) R(\omega,\gamma)\;\left (\langle\omega_s^2(\gamma)\rangle
-\langle\omega_s(\gamma)\rangle^2\right )\;,
\eqno(\new)
$$
where $\langle \omega_s^2(\gamma)\rangle$ is the second moment of
the distribution of scattered photons in the scattering of
the photon $\omega$ and electron $\gamma$.

We use the full kinetic equation
when the photon energy is $\omega> \omega_b(\gamma)$.  In this case, there is
no problem with the finite energy bin size because the change in photon 
energy when these higher frequency photons scatter is large and
$\Delta\gamma\sim (\gamma-1)$.
Equation (\compt) is then written in the form 
$$
\Bigl ( {\partial f(\gamma,t)\over \partial t}
\Bigr )_{e\gamma} =
-\,{\partial\over \partial \gamma}[\dot \gamma_C f(\gamma,t)]
+{1\over 2}\; {\partial^2\over \partial \gamma^2}\;[D_C(\gamma)
\, f(\gamma,t)]
-f(\gamma)\int\limits_{\omega_b(\gamma)}^{\omega_{max}}
 d\omega N(\omega) R(\omega,\gamma)
$$
$$\eqnam{\newcompt}
+  \int\limits_{\omega_b(\gamma)}^{\omega_{max}}
 d\omega' \int d\gamma' R(\omega',\gamma')
P(\gamma;\gamma',\omega') N(\omega') f(\gamma')\;,
\eqno(\new)
$$
where $\omega_{max}$ is the maximum photon energy in the spectrum.
To compute both $\gamma_C(\gamma)$ and $D_C(\gamma)$ numerically we
use a much finer resolution in photon energy space than that for
$\omega > \omega_b(\gamma)$.

We have tested the validity of this approach by first varying the `break
frequency' $\omega_b(\gamma)$ and found that the variations 
in the results are always much smaller than the chosen relative 
variation in $\omega_b(\gamma)$.  We also varied the number of bins $N$
for a fixed electron energy range and found that this too makes
a negligible difference on the results.  These tests would unambiguously
point to inconsistencies if they were there.

Finally, for pair annihilation we use the rate
$R(\gamma_{-},\gamma_{+})$ given by Svensson (1982), 
where $\gamma_-$ and $\gamma_+$ are electron and positron
gamma-factors, respectively. For pair
creation we adopt the expressions in Coppi \& Blandford (1990).

\bigskip
\centerline{\bf 6. Tests}
\medskip
\centerline{\sl 6.1. The Full Equation}
\medskip
Our goal in this section is to demonstrate the use of the 
self-consistent FP approach in the treatment of photon-particle interactions 
in astrophysical plasmas.  Situations where one can assume that a part of
the distribution is Maxwellian have been studied in some detail for the
case of non-thermal models (e.g,  Zdziarski et al. 1990;
Svensson 1994, and references cited therein). Ghisellini, Guilbert \& Svensson
(1988) studied the thermalization of the low energy end of the non-thermal
distribution by synchrotron reabsorption; they neglected e-e Coulomb
and inverse Compton interactions.  DL89 developed a perturbative approach 
to the FP equation for e-e Coulomb interactions, but have not applied 
the method to problems that require finding the photon distribution
self-consistently. 

We first present the complete equation, including all the processes discussed 
above, i.e., Coulomb and Compton interactions, as well as pair creation and 
annihilation. We shall not here attempt to make
a detailed study of the various applications, which would require a careful
examination of source geometries and a thorough search in parameter space.
This subject is vast and will be covered in future publications.

All the distributions are assumed to be spatially uniform inside of a 
spherical volume with radius $R$. Since the number of positrons is always 
different (i.e., smaller) than the number of electrons, the positron 
distribution is never identical to that of the electrons.  This is true 
even though the electron-electron and electron-positron cross sections
are the same within the scope of the approximation we use here (see
\S 4.1).  The processes that shape these particle profiles differently
include pair annihilation and, when included, bremsstrahlung emission 
and pair escape.
In the following, we neglect the contribution of bremsstrahlung
to the overall photon production, assuming instead that the soft photons 
are injected by an external source.  Index $1$ will be used to denote
electrons, whereas $2$ pertains to positrons.

The full equation for the electrons reads
$$
\Bigl ( {\partial f_1(\gamma,t)\over \partial t}
\Bigr ) =
-\,{\partial \over \partial E }
\Bigl [ A(\gamma) f_1(\gamma,t) \Bigr ]\,+\,
{1\over 2}\,{\partial^2 \over \partial E^2 }
\Bigl [ D(\gamma) f_1(\gamma,t) \Bigr ]\,+
$$
\smallskip
$$\eqnam{\grand}
+\,\,\Bigl ( {\partial f_1(\gamma,t)\over \partial t}
\Bigr )_{e\gamma}\, - \,R_1(\gamma)\, f_1(\gamma,t)
+\, S(\gamma)\;,
\eqno(\new)
$$
where $A(\gamma)$ is the total energy exchange  rate for an electron
with Lorentz factor $\gamma$; that is, $A(\gamma)$ is the sum of
energy exchange rates from processes which can be
treated using a Fokker-Planck approach.  This excludes inverse Compton
interactions which are instead incorporated into the term
$({\partial f_1(\gamma,t)/ \partial t} )_{e\gamma}$,
as given in \S 4.2. 
Similarly, $D(\gamma)$ is the electron diffusion 
coefficient, a sum of the electron-electron, positron-electron and 
proton-electron diffusion coefficients.

The function $R_1(\gamma)$ is the electron annihilation rate,
$$\eqnam{\ynicta}
R_1(\gamma)= \int\limits_{1}^{\infty} d\gamma_2 f_2(\gamma_2)
R(\gamma,\gamma_2)\;,
\eqno(\new)
$$
where $R(\gamma,\gamma_2)$ is the angle-averaged annihilation rate
for an electron with energy $\gamma$ and positron with energy
$\gamma_2$ as given by Svensson (1982).
$S(\gamma)$ is a `source' term that represents 
pair production by photon-photon collisions.
The pair creation rate is an integral over the photon distribution 
$N(\omega,t)$ as given by Coppi \& Blandford (1990).

The full equation for the positrons has an identical form,
except that the subscripts $1$ and $2$ are interchanged.
The often used quantitative measure of the heating power going into
electrons is the 
dimensionless `hard' compactness parameter
$$\eqnam{\lh}
l_h = \,{4 \pi R^2 \sigma_T \over 3 c}
\int\, d\gamma \;A^+(\gamma,t)\;[f_1(\gamma,t) + f_2(\gamma, t)],
\eqno(\new)
$$
where $A^+(\gamma,t)$ is the energy and  time dependent
heating rate for an electron/positron with energy $\gamma$. For the 
proton heating rate given by Equation (\arbit), the time dependence 
results from the change of the proton temperature in time. Letting
$l_h$ represent the energy transfer rate to the protons
(due to dissipational processes), the proton temperature
is then found by balancing this heating and 
the Coulomb cooling by cold electrons at every instant in time.

The equation for the photons reads
$$\eqnam{\photeq}
{\partial N(\omega)\over \partial t}=
\dot N(\omega) - \Bigl ({\partial N(\omega)\over \partial t}\Bigr )_{e\gamma}
+ P(\omega) - N(\omega)/t_{esc}(\omega)\;,
\eqno(\new)
$$
where $\dot N(\omega)$ is the rate of external photon injection
(photons are here assumed to be injected uniformly in space throughout
the plasma containing region), $P(\omega)$ is the number of photons with 
energy $\omega$ created per unit time per unit
volume per unit energy by pair annihilation, and $t_{esc}
(\omega)$ is the photon escape time as given by Lightman \& Zdziarski 1987:
$$\eqnam{\phesc}
t_{esc}(\omega) = {R\over c}\; [1+ (1/3)\,
\phi(\omega)\;\tau_{\rm KN}(\omega)]\;,
\eqno(\new)
$$
in terms of the size $R$ of the region. The energy dependent optical thickness 
$\tau_{\rm KN}(\omega)$ is here defined by 
$$
\tau(\omega) = \tau_T \sigma_{\rm KN}(\omega)/\sigma_T\;,
$$
where $\tau_{\rm KN}(\omega)$ is the Klein-Nishina cross-section (see, e.g.,
Jauch and Rohrlich 1980, Eq.[11.24]). The function $\phi(\omega)$ 
takes into account the fact that forward scattering for high energy 
photons becomes important (see Lightman and Zdziarski 1987; Equation [21b]).
We should mention that this treatment of the photon radiative transport is
only approximate, especially for a low optical depth. However, the
solution found this way will undoubtedly bear the same qualitative imprints
of the `exact' electron distribution as the still to be found 
solution which includes {\it both} radiative transport 
for the photons and the FP treatment of the
electrons.

The term $[\partial N(\omega)/
{\partial t}]_{e\gamma}$ represents the full Compton interaction with 
electrons and positrons and is given by the collision integral
$$
- \Bigl ({\partial N(\omega)\over \partial t}\Bigr )_{e\gamma}=
-N(\omega)\int\limits_{0}^{\infty} d\omega 
\int\limits_{\,1}^{\infty} d\gamma R(\omega,\gamma)\,
[ \,f_1(\gamma) + f_2(\gamma)\, ] 
$$
$$\eqnam{\kompfot}
+\int\limits_{0}^{\infty} d\omega' 
\int\limits_{\,1}^{\infty} d\gamma' P(\gamma,\gamma',\omega')
R(\omega',\gamma')N(\omega')\,[ f_1(\gamma') + f_2(\gamma') ]\;,
\eqno(\new)
$$
where $R(\omega,\gamma)$ and $P(\gamma,\gamma',\omega')$
are defined in \S 4.2, and $\gamma+\omega = \gamma'+\omega'$.

In the tests reported here, we take the photon injection 
rate $\dot N(\omega)$ to be that of a blackbody spectrum with dimensionless
temperature $\Theta_b \equiv k T_b/m_ec^2$ in the range 
$\Theta_b = 10^{-5} - 10^{-4}$. 
This injection rate is characterized by the dimensionless `soft' photon 
compactness parameter (e.g., Lightman \& Zdziarski 1987)
$$\eqnam{\softl}
l_s = {4 \pi R^2 \sigma_T \over 3 c}
\int d\omega \,\omega \,\dot N(\omega)\;.
\eqno(\new)
$$

We also need to specify the electron heating mechanism. 
It is well known (Guilbert, Fabian \& Stepney 1982), that the 
electron-proton Coulomb interactions are very inefficient,
and if cooling is effective, it is
impossible to account for the observed high
electron temperature $\Theta_e\simgt 0.2$ observed in Seyfert Galaxies,
as implied by their high exponential cutoff energy 
(e.g., Svensson 1996 and references therein).
One then needs to introduce some {\it ad hoc} heating mechanism for the
electrons. Many workers do not specify the mechanism by which
the energy is transferred to the electrons and instead specify the
heating rate per electron (independent of the electron energy)
per unit volume. This is indeed the only information needed if one 
assumes that the particles are thermal.

However, for our purposes we need
to know the diffusion coefficient for this heating interaction.
Since it is beyond the scope of this paper to
explore the full range of interactions that heat the electrons in
real astrophysical plasmas, we
model the electron heating by the e-p Coulomb 
interactions with a fixed proton temperature,
$kT_0 = 20$ MeV,  but in addition,
we multiply this interaction (both the energy exchange and diffusion) 
by a factor (equal to $T_p/T_0$, see \S\  6.4.c) 
which is needed to account for the electron
heating rate given by the specified $l_h$. While this is 
obviously a non-physical interaction, it will allow us to make some
progress in understanding what one can expect to happen with the
electron distribution. It also will allow us to have a `reference scale';
 whether the electron distribution is
actually broader or narrower than the one found in this test model
depends on whether the actual electron heating mechanism produces 
comparatively more or less diffusion. 
\bigskip
\centerline {\sl 6.3. Results}
\medskip
In this section we apply our technique to a test situation where
the electrons and positrons are heated by some external source 
as described in section 6.1, and cooled by inverse 
Compton scatterings with soft intense external radiation. 
The description of the numerical procedure used to solve
for the exact electron distribution can be found in Appendix A.3.
Ghisellini, Haardt and Fabian (1993) (GHF hereafter)
have shown that the exact shape of the electron distribution produces very
little effect on the photon spectra as long as the { \it multiple} Compton
scatterings are the main emission mechanism. In particular, they considered
three different particle distributions that had the same normalization and
the same value of $\langle {\gamma^2-1}\rangle$: 
a Maxwellian, a power-law, and a constant
(i.e., $f(\gamma) = $const). The resulting spectra were hardly distinguishable.

We conduct similar tests here. For the test electron distribution, we choose:
1) the exact electron distribution found by using the techniques
developed in this paper;
2) a Maxwellian; 3) a  distribution broader than a
Maxwellian (see Figs. 11a and 12a).
 In the last case, the functional shape of
the distribution function  is given by
$f(E) = {\rm const}\,\sqrt{f_{\rm mxw}(E,\Theta_e)}$, 
where $f_{\rm mxw}(E,\Theta_e)$
is a Maxwellian distribution with temperature $\Theta_e$. 
Just like in the familiar Maxwellian case, there are again two
parameters, $\Theta_e$ and the normalization, which are found self-consistently.
 One very important difference of our tests from those of
GHF is that 
we find the equilibrium pair number density 
by solving the exact pair balance,
so that the relative positron number density, $z\equiv n_+/n_p$, is
different for the three tests.
 This results in a different optical depth for 
the different distributions, even though the proton optical depth
 $\tau_p$ is  the same for all three.

We first discuss the photon spectra. 
Figures 11b and 12b show the photon spectra 
for $l_h=l_s = 420$, $\tau_p =0.05$ and $l_h =8.4 $, $l_s = 2.1$,
$\tau_p = 0.02$ respectively.
The input and output parameters of these tests are given in Table 1.
For the case of the higher optical depth (Fig. 11b),
it is readily seen that the spectrum
does not vary much from one case to another, at least in the X-ray energy
range (2-18 keV).
This is similar to the findings of GHF, and is explained by the fact 
that the spectrum is formed by many repeated scatterings, which simply form
a power-law after two to three interactions. 

However, for the second test our
results disagree with conclusions of  GHF. We find 
qualitative differences between the  Comptonized spectra produced by the three
test electron distributions (Fig. 12b). The difference with GHF 
is caused by the fact that the calculated here exact electron distribution is 
narrower than a thermal one (Fig. 12a), and also that we 
test the parameter space with a lower optical depth. 
GHF considered only a subset of the parameter space available for
 the distribution function; they only considered cases where
the electron distribution is broader, more diffuse than the thermal 
distribution. At the present, it is completely 
unclear if this is indeed what happens. The answer to the question
regarding the electron distribution function can only be given if one 
specifies the interactions affecting  the electrons.

\medskip
\centerline{\sl 6.4. The Strongly Non-Maxwellian Electron Distribution}
\centerline{\sl a. The electron distribution profile}
\medskip

Using the tests described below as examples, we will discuss the 
parameter space where $\lambda$ is much larger than unity, i.e.,  
where strongly non-thermal distributions are expected.
Figures 11a and 12a show the three test electron distributions
for the tests described in 6.3 ($\lambda\simeq 80$ and $10$ for the
two figures, respectively).
The exact electron distributions are shown as  solid curves, and 
are much sharper than the thermal electron distribution (dashed line).
For the parameter space considered here, the electrons 
with  energy $E\ll\langle E\rangle$, where $\langle E\rangle$ is the 
electron average energy, interact mostly with protons which strive
to push these electrons to higher energy. The electrons with energy
 $E\gg\langle E\rangle$, similarly interact mostly with photons. 
Therefore, the electron distribution appears to be squeezed by the
photon cooling on the high energy end and the proton heating on the 
low energy end (see Fig. 13).

As surprising as this might sound, the exact distributions in Figures
11a) and 12a) are actually as simple to find as  the familiar
 Maxwellian distribution (corresponding to 
the given compactnesses and $\tau_p$). The FP part of the full electron
Equation (\grand) is by far the dominant one. That means that the actual
shape of the particle distributions is controlled by the FP part, while the
normalizations are governed by the much slower annihilation
and creation processes. Thus we can find the shape of the distributions
solving just the stationary FP equation. Furthermore,  for $\lambda\gg 1$,
both electron heating and cooling greatly exceed the e-e Coulomb interactions
(for an example see Fig. 13).
Essentially, the FP equation becomes linear in the electron distribution
function, which greatly simplifies the solution of this equation. 
The stationary (i.e., with zero time derivative) FP Equation (\fp) has then
an exact solution:
$$\eqnam{\vuola}
f(\gamma) = {{\rm const}\over D(\gamma)}\, \exp\left [2\int\limits_1^\gamma
d\gamma'\, [A(\gamma')/D(\gamma')]\right ].
\eqno(\new)
$$
We have dropped subscripts 1 and 2 on the distribution function since 
in this approximation the positron distribution has the same shape
as that of the electrons, and only the normalization is
different.\footnote{\hbox{$^{\dag}$}}{This similarity in the electron
and positron distributions exists in the exact simulations as well, and
we therefore do not show the positron curves in Figures 11a \& 12a.}

Figure 14 shows the exact distribution (same as in Fig. 12a)
and various fits to it. The heavy dashed curve shows the function given by 
Equation (\vuola) when $A(\gamma)$ and $D(\gamma)$ include all the interactions.
The larger dots show the same Equation (\vuola) but with the
e-e interactions excluded, i.e., the linear case. As one can see,
both fits are very good. 
And the faint dotted curve gives an analytic fit to 
Equation (\vuola).

We now explain this fit. Figure 13
shows all the FP coefficients for the test previously shown in Fig. 12.
Notice that the proton diffusion coefficient is almost a constant, so
we can approximate $D(\gamma)\simeq D(\gamma_0)\equiv D_0$ 
(where $\gamma_0$ is the 
point where electron cooling is equal to electron heating).
We  also approximate the electron-proton energy exchange coefficient
as a constant, $A^+(\gamma)\simeq A_0\equiv A^+(\gamma_0)$, which
should be sufficiently accurate around the peak energy $E_0$ (see the Figure).
Using  the fact that $A^-(\gamma)\propto
\beta^2\gamma^2$, we can write $A(\gamma)\simeq A_0 \bigl[1-(\beta\gamma/
\beta_0\gamma_0)^2\bigr ]$. Using this expression in (\vuola),
we get
$$\eqnam{\crude}
f(\gamma)\simeq {\rm const}\,\exp\left [ {2 A_0\over \beta_0^2\gamma_0^2
D_0}\,\bigr (\gamma_0^2\gamma - \gamma^3/3\bigr )\right ].
\eqno(\new)
$$
Notice that this expression is also quite accurate around the peak of the 
exact distribution function, but deviates on the low energy end. 
This deviation will be negligible for the distribution as a whole if
the electron heating is more or less energy independent.   

One can easily show that the dispersion of the
electron  distribution, defined as
$\langle{\Delta \gamma}\rangle \equiv \, 
\bigl [\langle(\gamma-\gamma_0)^2\rangle\bigr ]^{1/2}$,
where the averaging is taken over the electron distribution, is 
$$\eqnam{\dis}
\langle{\Delta \gamma}\rangle\;\simeq \;\left [ {\gamma_0\beta_0^2 D_0
\over A_0} \right ]^{1/2}
\eqno(\new)
$$
Therefore, for the same heating rate (which will lead to the same $\gamma_0$),
it is the diffusion coefficient that determines the dispersion of the 
distribution. Physical conditions under which the `wide' electron
distribution might form can now be understood in terms of a very large 
diffusion coefficient which might result from some particular
form of the electron interaction. 
\medskip
\centerline{\sl 6.4.b. Normalization of the electron distribution}
\medskip

The electron-positron annihilation rate depends rather weakly on
the energy of the colliding particles as long as their 
energy is smaller than a few MeV (e.g., Coppi and Blandford 1990,
Equation 3.7). Therefore, the exact particle profile does not
change the pair annihilation rate substantially. Nevertheless, 
its importance shows up in the shape of the photon spectra
 around the pair producing 
energy (see Figs. 11b and 12b). We can indirectly find the 
equilibrium pair number
density by considering the Compton $y$-parameter (see, e.g., Rybicki and
Lightman 1979). It is defined  as 
$$\eqnam{\ypar}
y\,\equiv\,\langle G \rangle\,\langle t_{\rm esc}\rangle/t_T,
\eqno(\new)
$$
where $\langle G\rangle$ is the average  fractional gain in the
photon energy in one scattering, $\langle t_{\rm esc}\rangle $
 is the average escape time and $t_T$  is the Thomson time.
By definition, the electron cooling rate is 
$\bigl (dE/dt)^- = c\sigma_T 
\langle G\rangle U_{\rm ph}$, where $U_{\rm ph}$
is the photon energy density in the source. In equilibrium, 
the total electron heating, $L_h$, equals the total cooling,
so that
$$\eqnam{\theat}
L_h\,=\, V_0 n_p (1 + 2z)\,(dE/dt)^- = c\sigma_T n_p (1+2z) \langle
G \rangle \,
V_0\, U_{\rm ph}. 
\eqno(\new)
$$
where $V_0$ is the source total volume and $z\equiv n_+/n_p$, 
ratio of positron number density to that of protons. Using Equation (\ypar) and 
the fact that  $L=U_{\rm ph} V_0/\langle t_{\rm esc}\rangle$, 
where $L = L_h + L_s$ is the total luminosity of the source, we get
$$\eqnam{\ygener}
y\,=\,{L_h\over L_h + L_s}.
\eqno(\new)
$$
This derivation did not make any use of the particular shape of the electron 
distribution, and therefore is valid for an arbitrary one, as long as 
the inverse Compton scattering is the main mechanism for the transfer of
energy from the particles to the escaping photons. We note that the numerical
value predicted by Equation (\ygener) should be corrected by taking into
account relativistic Klein-Nishina decline in the Compton cross section.
The principal effect then is to reduce $L_h$ in the denominator, since
these are the photons that have largest energies. This correction is rather
trivial to do, and we leave it to the reader.

Now, since we know the electron distribution shape (Eq. $\vuola$),
we can find $\langle G\rangle$ by simple convolution of this shape with
the particle-photon energy exchange rate. Equations (\ygener) and
(\ypar) then determine the Thomson optical depth  and so the desired normalization
of the electron distribution. This is particularly
simple for the Thomson scattering limit, e.g., for a spherical geometry, 
$$\eqnam{\yexpr}
y\simeq \tau_T (1 +\tau_T/3) (4/3) 
\langle\beta^2 \gamma^2\rangle.
\eqno(\new)
$$
For the tests shown in Figure 11, the $y$-parameter 
must equal $1/2$.  Numerically finding $\langle\beta^2 \gamma^2\rangle$ in 
Equation (\yexpr), we get $y=0.5\pm 0.01$ for all  three tests.

We have shown that, once the electron energy exchange and diffusion 
coefficients
 are known, it is a simple matter to find the electron distribution. 
Although we have considered only systems where the electron cooling is
dominated by the Comptonization of the soft external photons, one can
extend the methods described here to more complicated situations.
We intend to cover this in future publications. 

\centerline{\sl 6.4.c. Proton temperature}
\medskip
The last column in Table 1 shows the equilibrium proton 
`temperature'. We define the proton temperature such that the 
e-p heating is equal to $(T_p/T_0)\, a_p(\gamma,T_0)$, and $kT_0 = 20$
MeV. For both tests, we see that $T_p$ varies from distribution to
distribution, and the pattern is such that the more diffuse the 
distribution is, the colder the protons become. 

The explanation for this is rather straightforward. 
The e-p energy exchange rate 
(see Fig. 9 in this paper and Fig. 10 in DL89)
is a decreasing function of the electron  energy. It is mostly the low 
energy electrons that are heated by the e-p interactions.  Since more
diffuse distributions have more electrons at low energy, 
they absorb more energy per electron from the hot protons.  Additionally, 
we also saw in the tests we conducted that the more diffuse distributions 
are associated with the larger optical depths. That is, there are more 
leptons per proton, which again lowers the proton temperature.

Yet another consequence of the change in the electron 
distribution function is an associated change in the proton cooling 
time scale.  Let us define this time scale by the equality $E_{\rm p}/t_{\rm p}
\equiv A_{\rm p}$, where $A_{\rm p}$ is the proton cooling rate, and 
$E_{\rm p}$ is the proton average energy.
Assume that the proton heating rate (e.g., from the
dissipation of gravitational energy) is a constant. Then $A_{\rm p}$
is independent of the electron distribution, since proton cooling
and heating are equal to each other in equilibrium. Also, $E_{\rm p}
\propto T_{\rm p}$, the proton temperature, and thus 
the proton cooling time scale is directly proportional to this 
temperature. 

Even though we here assume a non-physical form for the proton heating,
it is quite likely that a similar result will hold for real
interactions that transfer energy from the hot protons to the
electrons in astrophysical plasmas. Indeed, the e-p energy exchange  rate
must be a decreasing function of the electron energy,
since after all it must become negative after the electron energy surpasses
that of the protons. The second part of the argument, based on the
change in lepton number, holds true irrespective of the exact
nature of the e-p interaction.  Finally, the proton cooling 
time scale will remain unchanged only when the proton-electron energy transfer
rate is independent of the proton energy, which is rather unlikely.

\medskip
\centerline{ \bf 8. Conclusions}
\medskip
We have presented the general FP method for finding the electron distribution
in non-Maxwellian hot, isotropic plasmas.
The technique can be applied to both time dependent and
stationary problems. Approximate expressions
(Equations $\vuola$ and $\crude$) for the exact electron distribution function 
are developed. These approximate expressions enable one to find the exact
electron distribution in about the same number of steps as is required to find
the temperature and equilibrium number density for a Maxwellian electron
distribution. 

A simple parameter (\S\ 2) specifies the section of the parameter space where the
treatment presented here is important. In general, this occurs for optically thin,
mildly relativistic and relativistic plasmas that are not too strongly magnetized.
Two representative tests 
show that the parameter space is further divided into two regions
(the exact boundary between these two regions depends on the
particle distribution and the as-yet-unknown physics of electron heating; 
we intend to investigate this boundary in future work).
In the first region the exact shape of the electron distribution does not 
influence the produced photon spectra, which are close to power-laws.
While this finding agrees with the work of GHF, we find that the difference in the
electron profile can lead to different equilibrium pair number densities.
This fact will be important in the radiation pressure dominated plasmas,
since the radiative pull on the protons is proportional to the number of
leptons per proton ($\equiv 1+2z$). Of equal importance is the change in the
proton temperature. A smaller proton temperature increases the importance of
the radiation pressure, and for the tests we have presented here, 
the combined effect is such that the ratio of radiation pressure to the
proton pressure changes by more than a factor of 10 from the exact to the `wide'
distribution. With such a large difference in this ratio, there
can be situations where the plasma can be either gas or radiation dominated 
depending on what distribution electrons have.
Therefore, we conclude that it  is probable that the equilibrium
(including hydrostatic) plasma state will still be sufficiently different
for different particle distributions to produce noticeable differences 
in the photon spectrum; this question needs to be investigated further.

In the other, generally optically thinner, region, the photon spectra
bear strong imprints of the exact shape of the particle distribution,
and are not simple power-laws. 
The local X-ray index and break energies depend on the exact
particle profile. Use of the exact electron distribution can then
limit the parameter space for a particular astrophysical model
differently than the use of the thermal distribution does. 

A shortcoming of this paper is the use of the diffusive escape time
approximation (Equations \photeq, \phesc) instead of the exact 
radiation transport 
approach (e.g., Poutanen and Svensson 1996). One can, however,  include
radiation transport in the same way as is done  for
Maxwellian electrons. We anticipate that the `non-power-law' features in the
photon spectra may be even stronger with the inclusion of radiation transport.
Our treatment of radiation employed the angle averaged corrected Jones
expression for the Compton scattering. If one is fixing the viewing angle instead,
the amount of dispersion in the photon upscattered energies become smaller 
(because one angle average is removed),
and the spectra will track the electron distribution better.

Another interesting question, which should be addressed in future work, is 
whether there are generic limits on the ratio of the electron heating diffusion
coefficient to the energy exchange one, independent of the exact interaction
which heats the electrons. If this is so, then we should be able to 
determine just how much the exact electron distribution can be broader or 
narrower than a Maxwellian.

\vskip 2 cm
C. Dermer provided us with his own differencing scheme
which helped us to developed the scheme presented in this paper.
We are grateful to M. Boettcher for a stimulating discussion of the
importance of the different terms in the electron-electron diffusion 
coefficient.  We are also grateful to the anonymous referee for his
careful reading of the manuscript and making many suggestions that 
have greatly improved the paper.
This work was supported in part by NASA grants NAGW-2709 and NAG 5-3075.
\vfill\eject\null

\centerline{\bf Figure Captions}
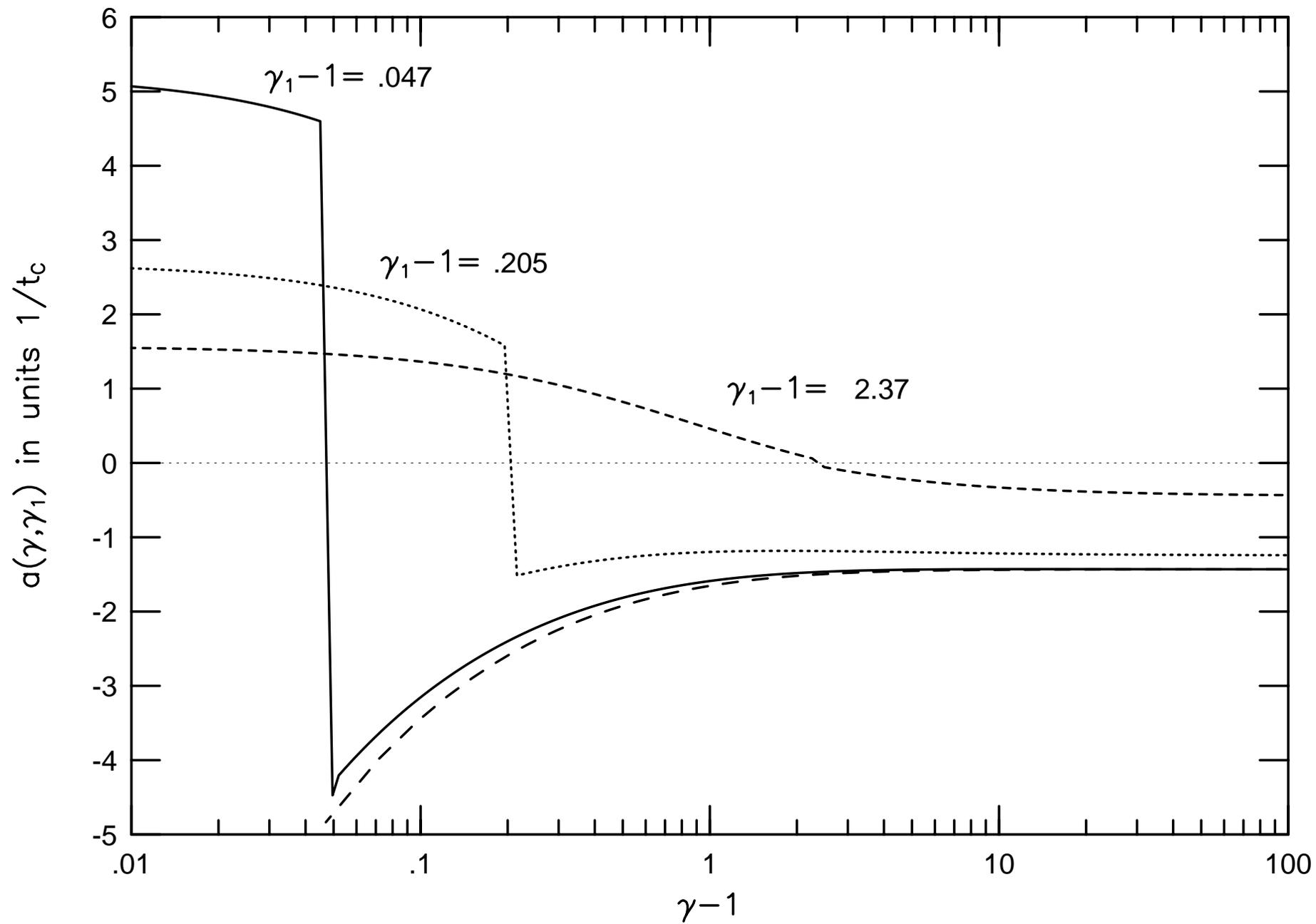
\noindent{\bf Figure 1. }
Coulomb energy exchange coefficient $a(\gamma,\gamma_1)$ 
for a test particle scattering off a mono-energetic electron distribution
(i.e., $f(\gamma) = \delta[\gamma-\gamma_1]$) for three different values of 
$\gamma_1$.  The curves are labeled by their 
corresponding values of $\gamma_1$. The Coulomb logarithm here and 
throughout the paper is set to 20. The time $t_C$
is defined as $t_C \equiv (n_e c \sigma_T \ln \Lambda)^{-1}$, and
$n_e$ is the total lepton number density.

\vskip 0.2in
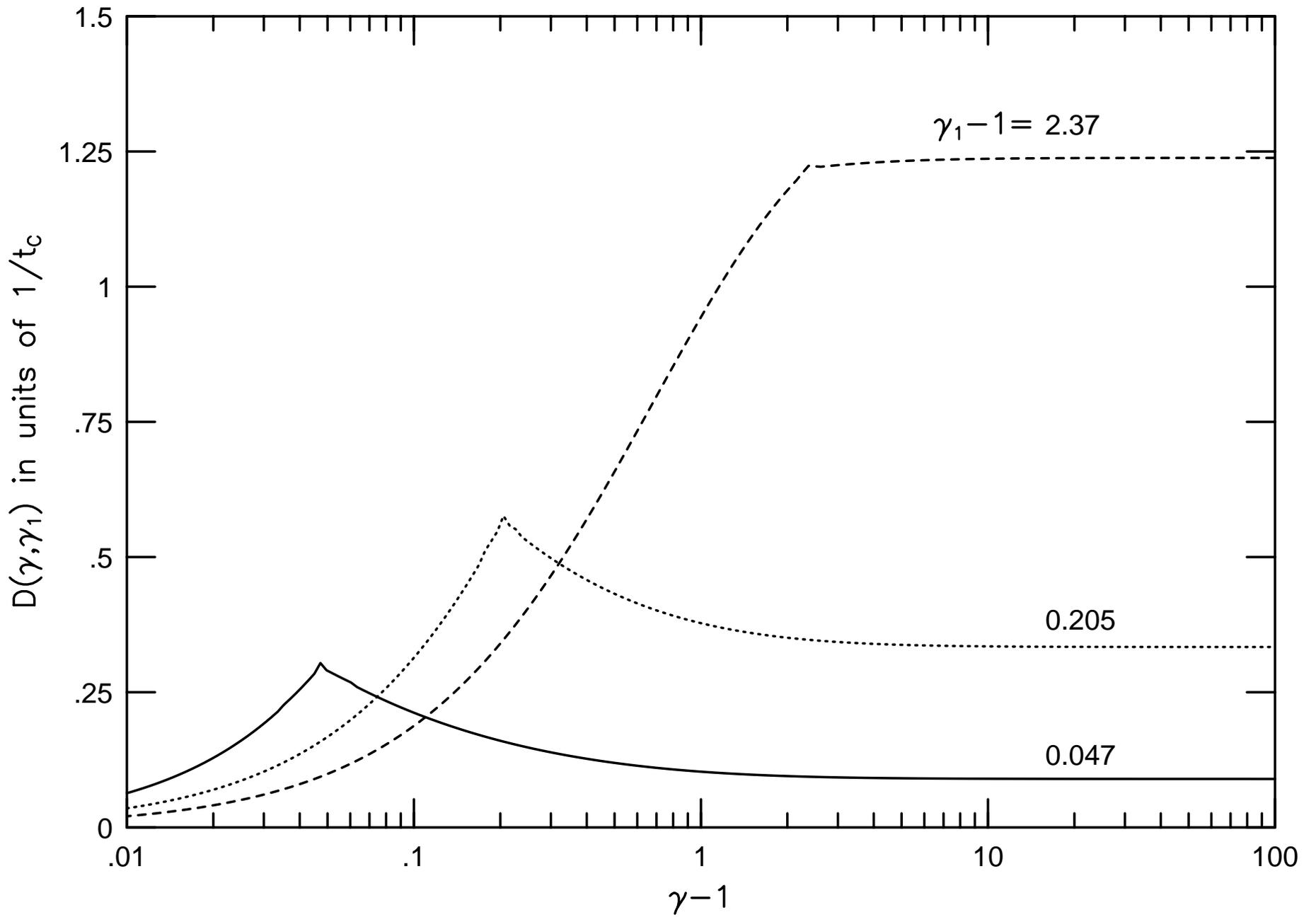
\noindent{\bf Figure 2.} Coulomb diffusion  coefficient $D(\gamma,\gamma_1)$
for a test particle scattering off a mono-energetic electron distribution
(i.e., $f(\gamma) = \delta[\gamma-\gamma_1]$) for three different values of 
$\gamma_1$.  The curves are labeled by their 
corresponding values of $\gamma_1$.

\vskip 0.2in
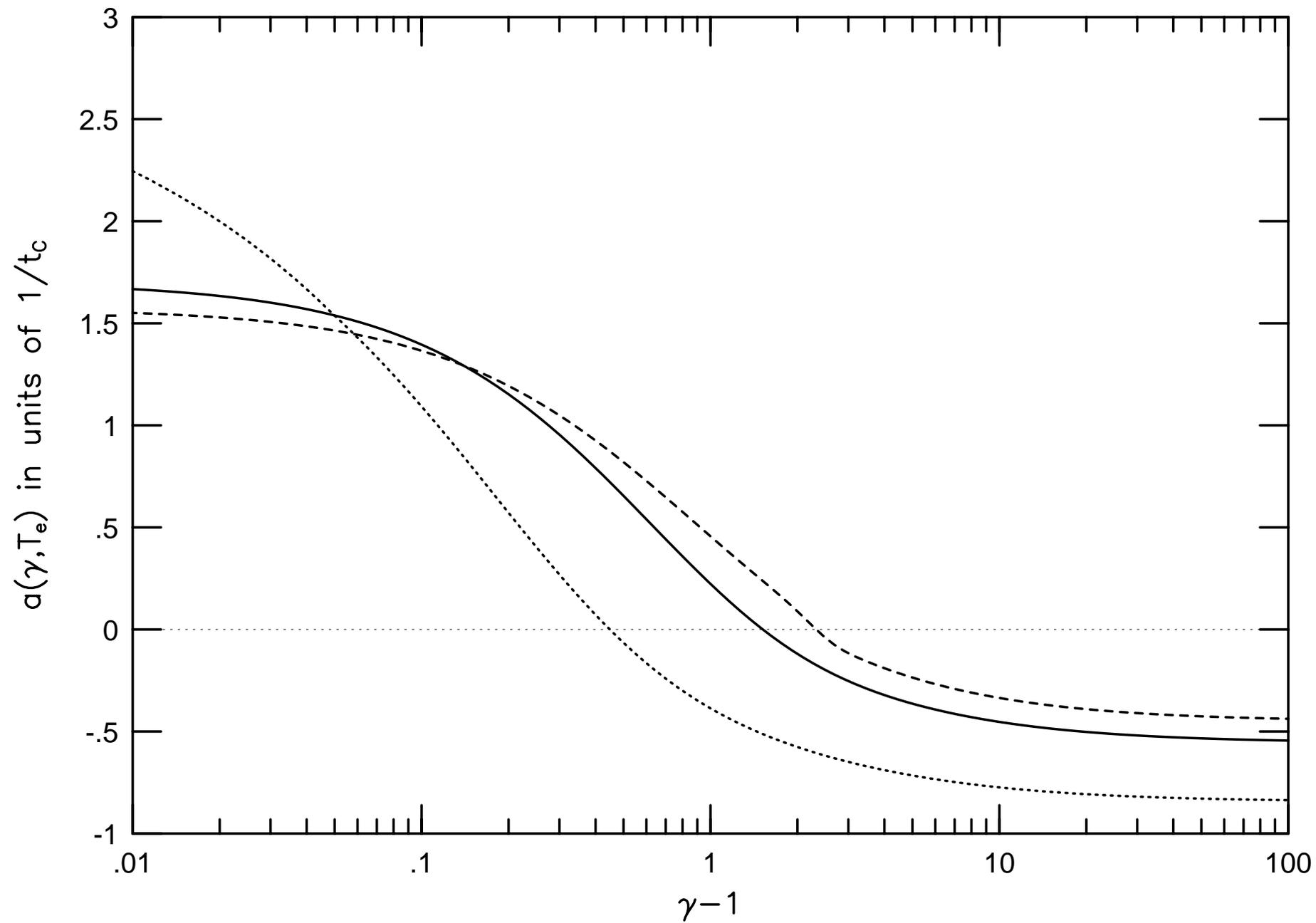
\noindent{\bf Figure 3.} The electron energy exchange coefficients $a(\gamma)$ 
for three different electron distributions (plotted in Fig. 5 below). The 
solid curve corresponds to a Maxwellian distribution with temperature
$\Theta_e\equiv kT_e/{m_e c^2} = 1$. 
 The distributions have the same average energy and 
are normalized to unity. The dotted and dashed curves correspond to
a power-law ($f(\gamma)=\beta\gamma^{-p}$; $p$ = 2.48) 
and a Gaussian, respectively. 

\vskip 0.2in
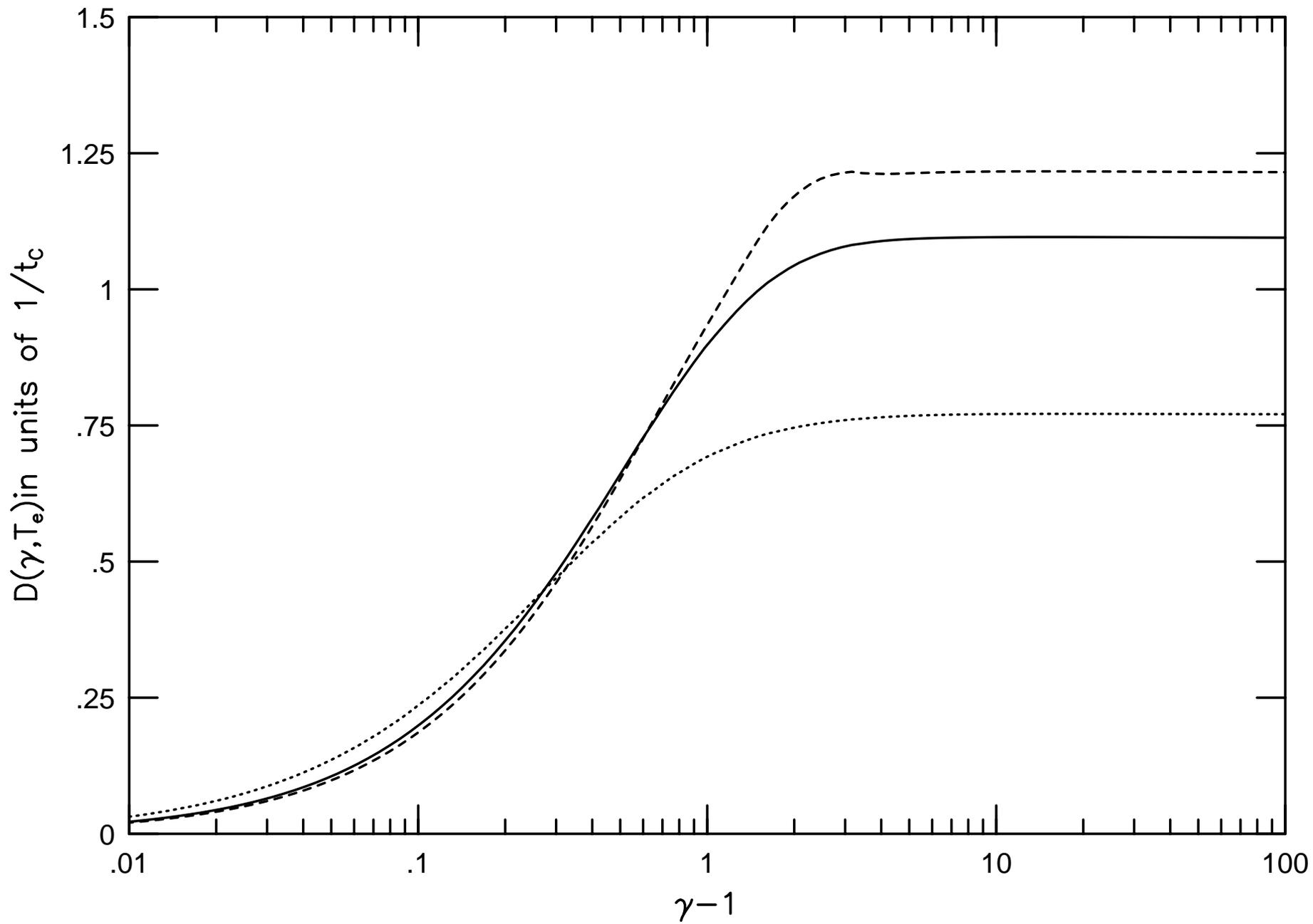
\noindent{\bf Figure 4.} The electron  diffusion coefficient $D(\gamma)$
for the same distributions as in Figure~3.

\vskip 0.2in
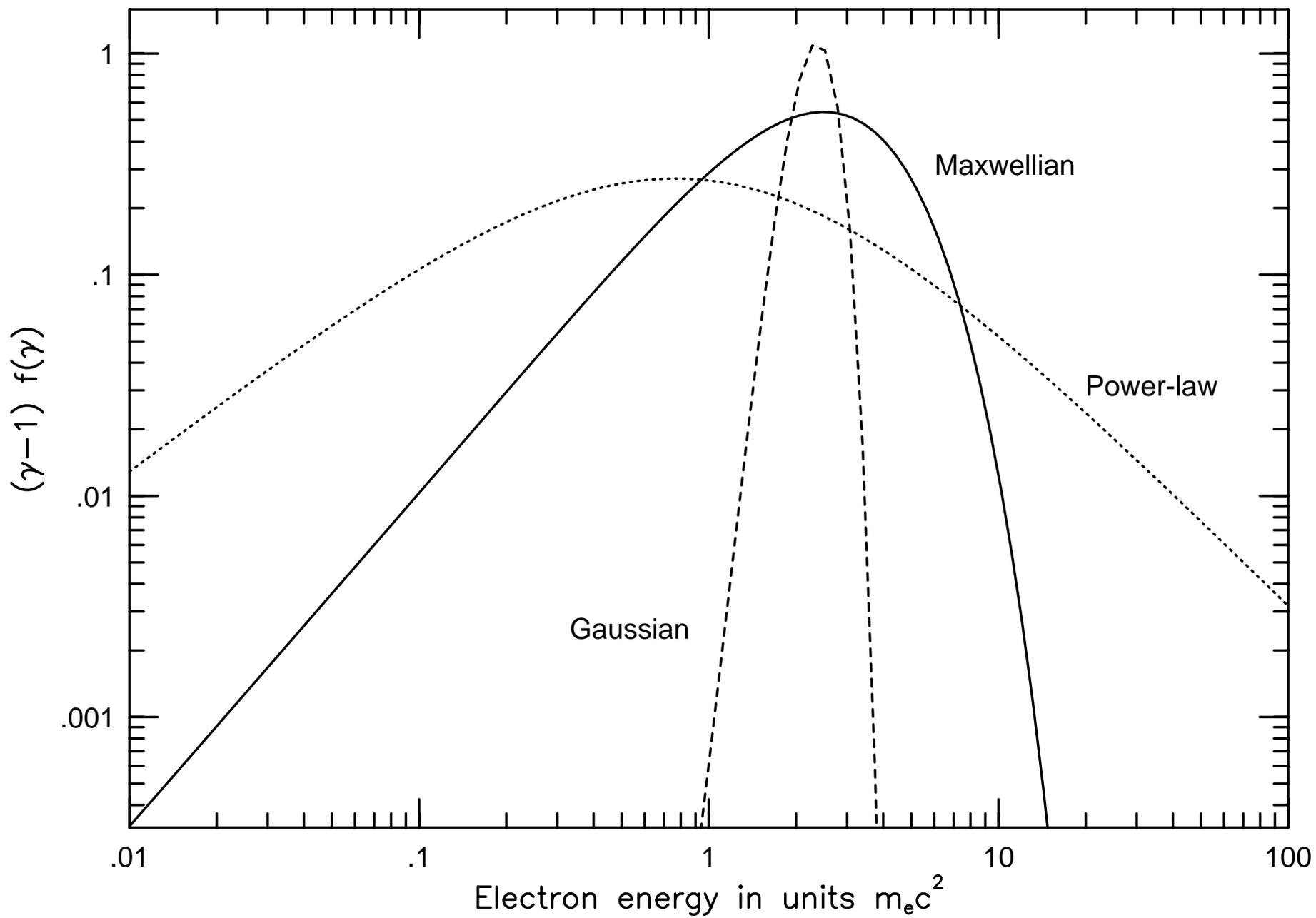
\noindent{\bf Figure 5. } The three electron distributions used to compute
the FP coefficients plotted in Figures 3 $\&$ 4.  All three functions are 
normalized to unity and have the same average energy. Solid, dotted and 
dashed lines correspond to Maxwellian, power-law and Gaussian distributions,
respectively.  The temperature of the Maxwellian is $\Theta_e = 1$.

\vskip 0.2in
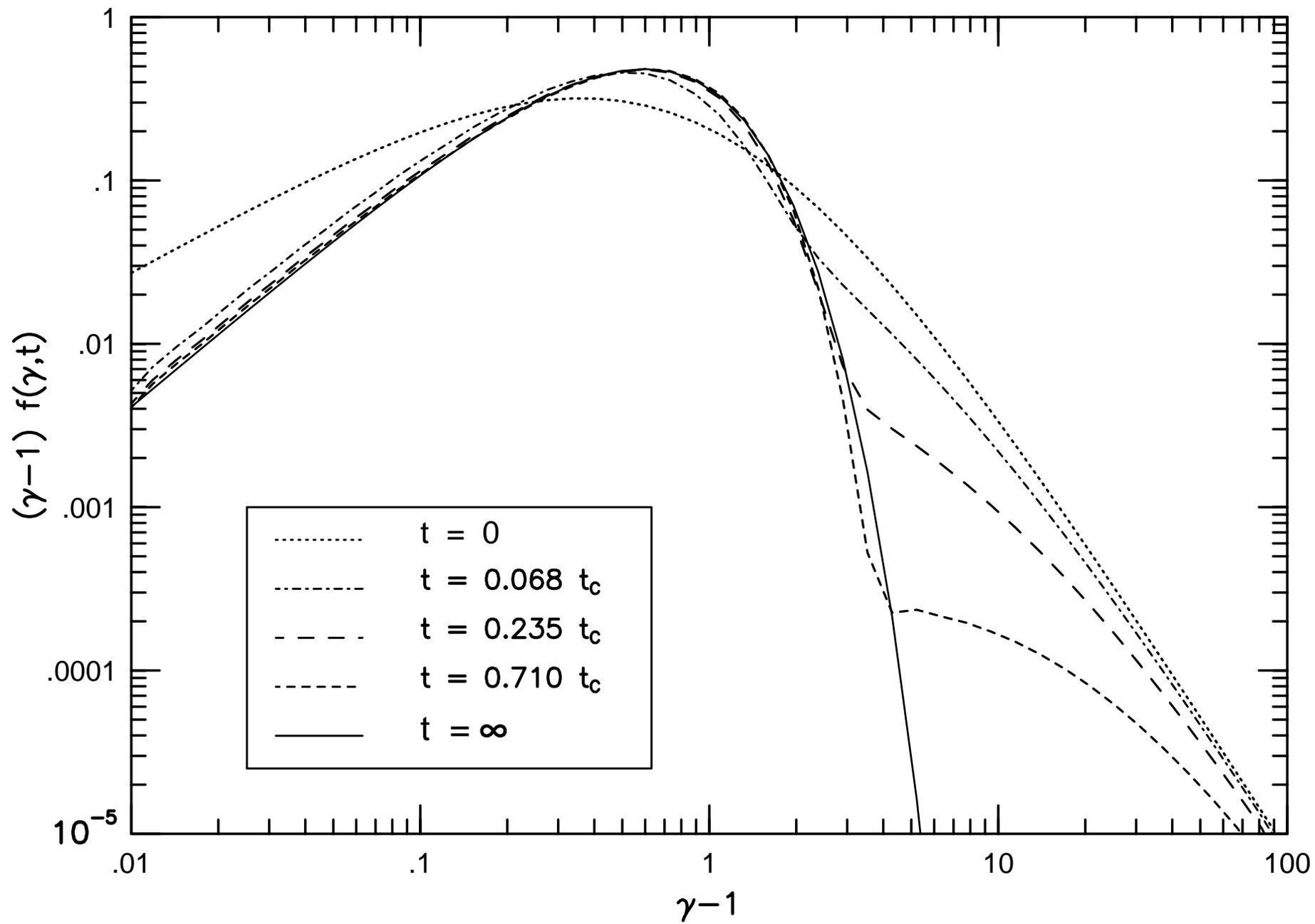
\noindent {\bf Figure 6.}
Time evolution of an initial power-law 
($ f(\gamma)=\gamma^{-p}$; $p$ = 3.72) distribution of electrons
coming into equilibrium (i.e., approaching a Maxwellian distribution) via 
Coulomb interactions with itself.
The temperature of the equilibrium Maxwellian is 0.3.
The box on the left shows the time corresponding to each curve
($t_C$ is defined in the caption of Fig. 1). The solid curve with $t=\infty$
corresponds to the perfect Maxwellian.
 Note that our
choice of times to plot $f(\gamma,t)$ was dictated by convenience of
viewing and not by equal spacing in time. In fact, thermalization
is always {\it slower} at {\it later} times.

\vskip 0.2in
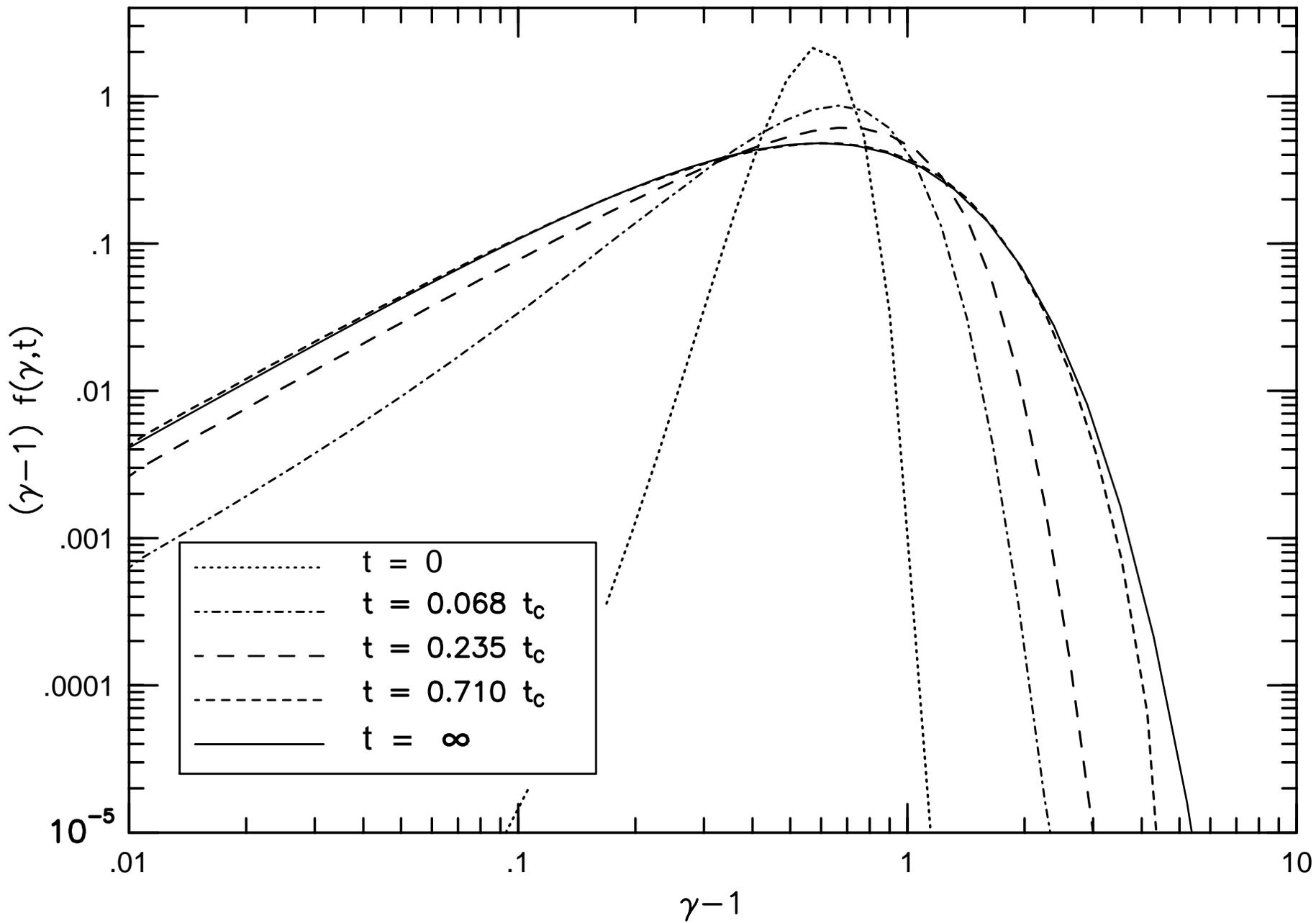
\noindent {\bf Figure 7.} The same as Figure 6 except that the initial
distribution is a Gaussian with the mean energy corresponding to a
Maxwellian temperature of 0.3 $m_e c^2$.

\vskip 0.2in
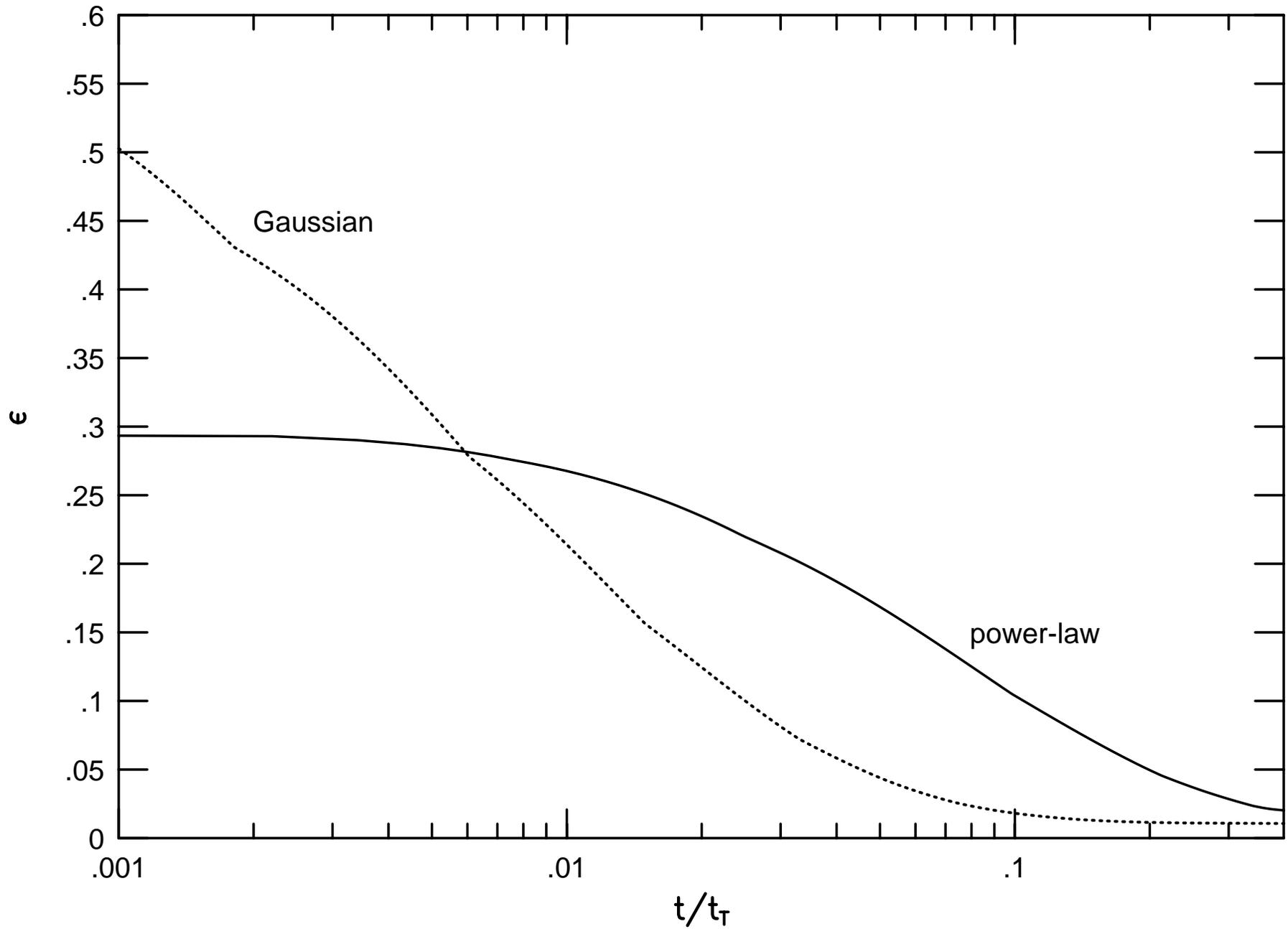
\noindent {\bf Figure 8.} Time evolution of 
$\varepsilon$ (defined in Eq. \deven), the deviation from a perfect
Maxwellian. Solid and dotted lines correspond
to thermalization of the power-law (Fig. 6) and Gaussian (Fig. 7).
Note the Gaussian relaxes to the thermal distribution 
much faster than the power-law. For comparison, the electron-electron 
thermalization time scale given by Equation (\stepney) is 0.06$\,t_{\rm T}$
for the given equilibrium temperature of 0.3.

\vskip 0.2in
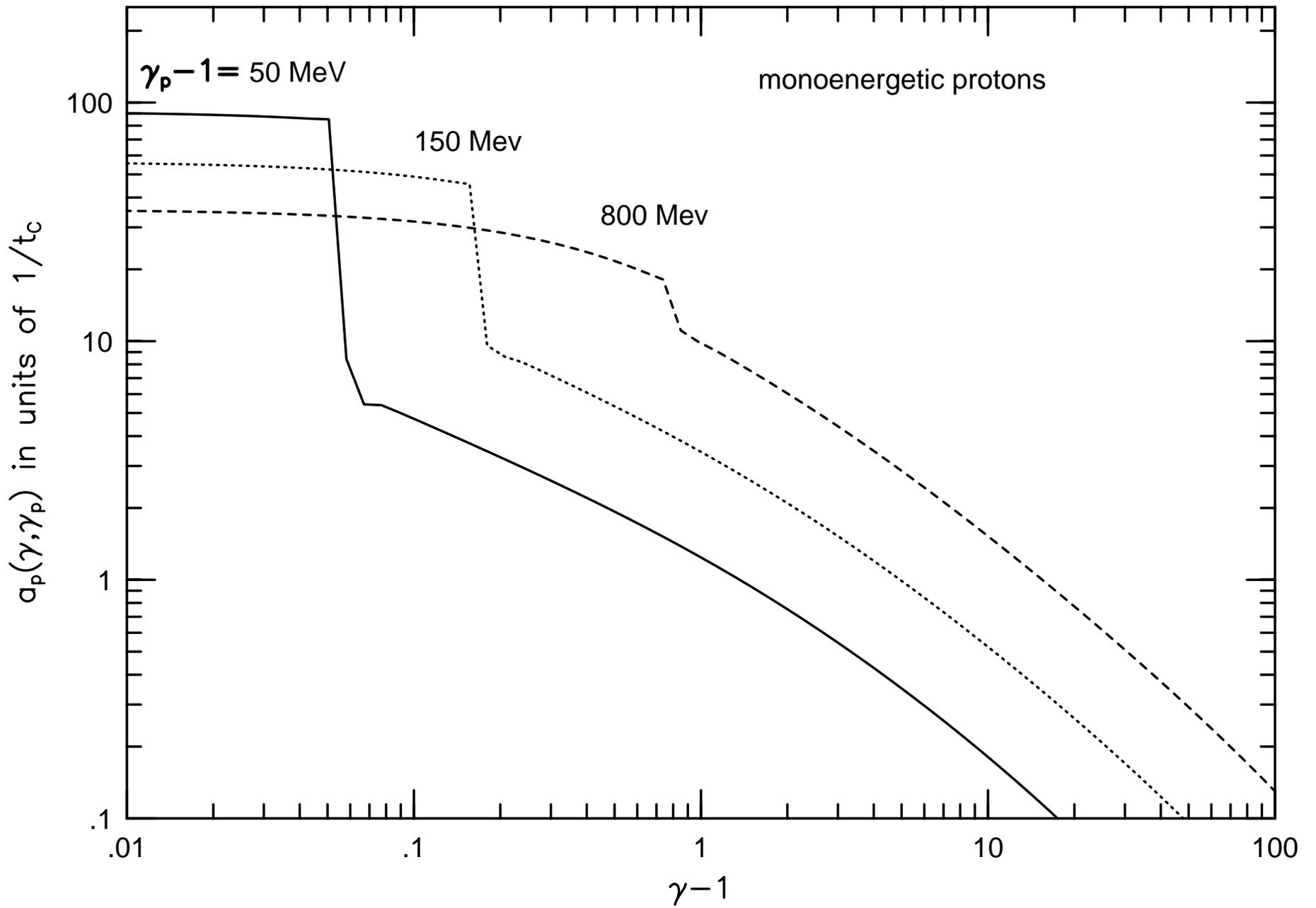
\noindent {\bf Figure 9.} The mono-energetic electron-proton energy
exchange coefficient $a_p(\gamma,\gamma_p)$ as a function of the electron
energy. The curves are labeled by their corresponding values of 
the proton kinetic energy.

\vskip 0.2in
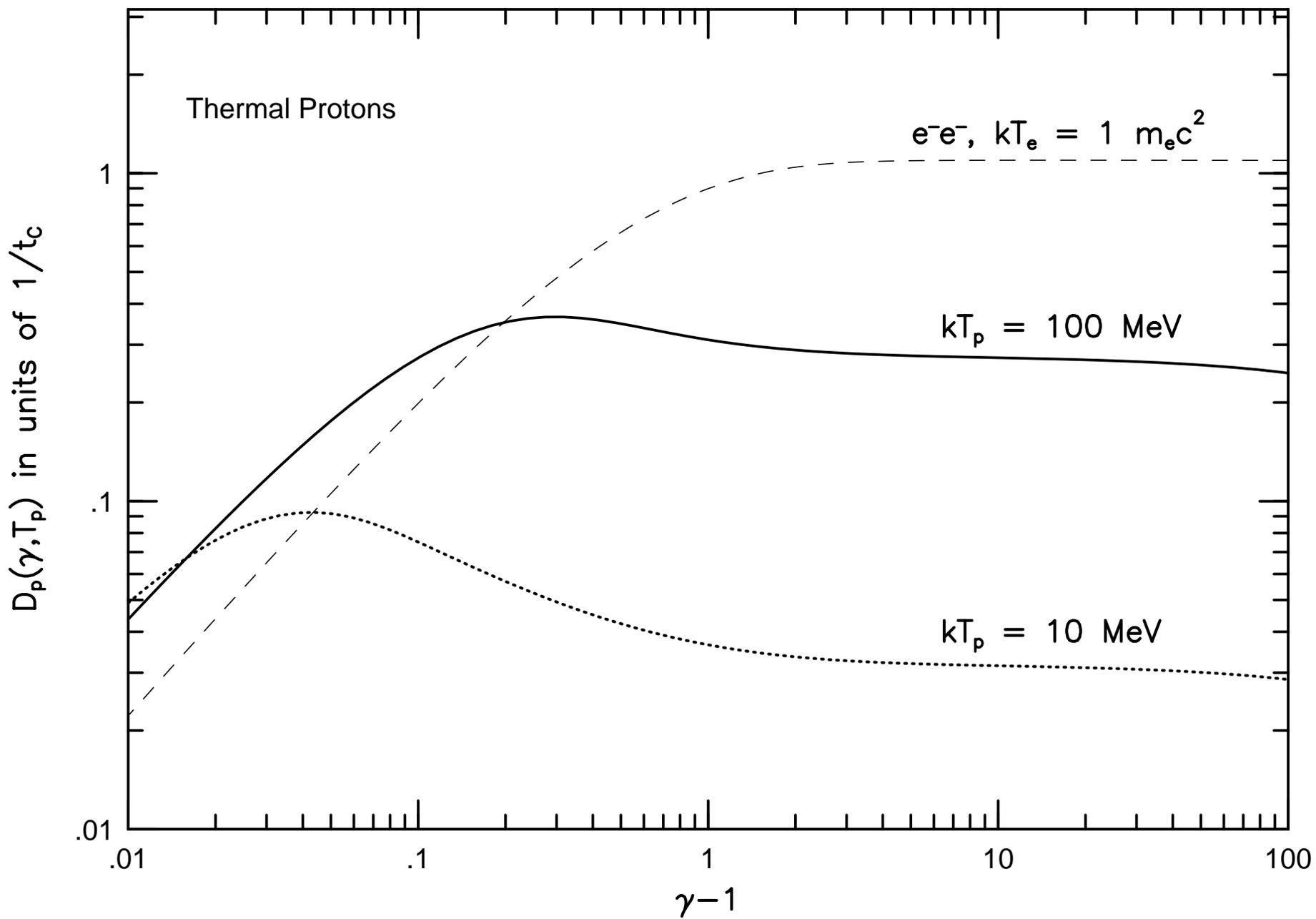
\noindent {\bf Figure 10.} The electron-proton 
diffusion  coefficient for
the case of Maxwellian protons. The Maxwellian electron-electron diffusion 
coefficient for $\Theta_e = 1 $ is shown 
with a dashed line for comparison (same as the solid curve in Fig. 4).

\vskip 0.2in
\noindent {\bf Figure 11.} {\bf a)} The equilibrium electron distributions
for the tests described in section 6.3.  The parameters are:
$l_h=l_s=420$, $\tau_p=0.05$ (also see Table 1).
The exact distribution is
the solution of the full kinetic equation, which includes electron 
heating, cooling and pair creation and annihilation.
The Maxwellian curve is the solution
of the same problem but assuming that the distribution is thermal. The
`wide' electron distribution is again a solution of the same problem,
but assuming a wider than thermal profile for the distribution.
The three distributions have a different `temperature' and number of particles,
but produce almost identical spectra (Fig. b). {\bf b)} The photon spectra
corresponding to the electron distributions plotted in Fig. 11a.

\vskip 0.2in
\noindent {\bf Figure 12.} {\bf a)} The same as Fig. 11a, but now for
$l_h=8.4$, $l_s =2.1$ , $\tau_p=0.02$. 
{\bf b)} The corresponding photon spectra. Notice that the spectra are distinctly
different from each other, and are not simple power-laws anymore.
See Table 1 for additional information about input and output parameters.

\vskip 0.2in
\noindent {\bf Figure 13.} {\bf a)} The electron energy exchange coefficients
plotted for the tests shown in Fig. 12.
The solid, dotted and dashed  curves show 
the electron-photon cooling (the dominant FP part, see \S\ 4.2),
e-p heating and e-e energy exchange coefficients. 
{\bf b)} Same as {\bf a)}, but for the electron diffusion coefficients.
Note that both energy exchange and diffusion are dominated by 
electron-photon or electron-proton interactions.

\vskip 0.2in
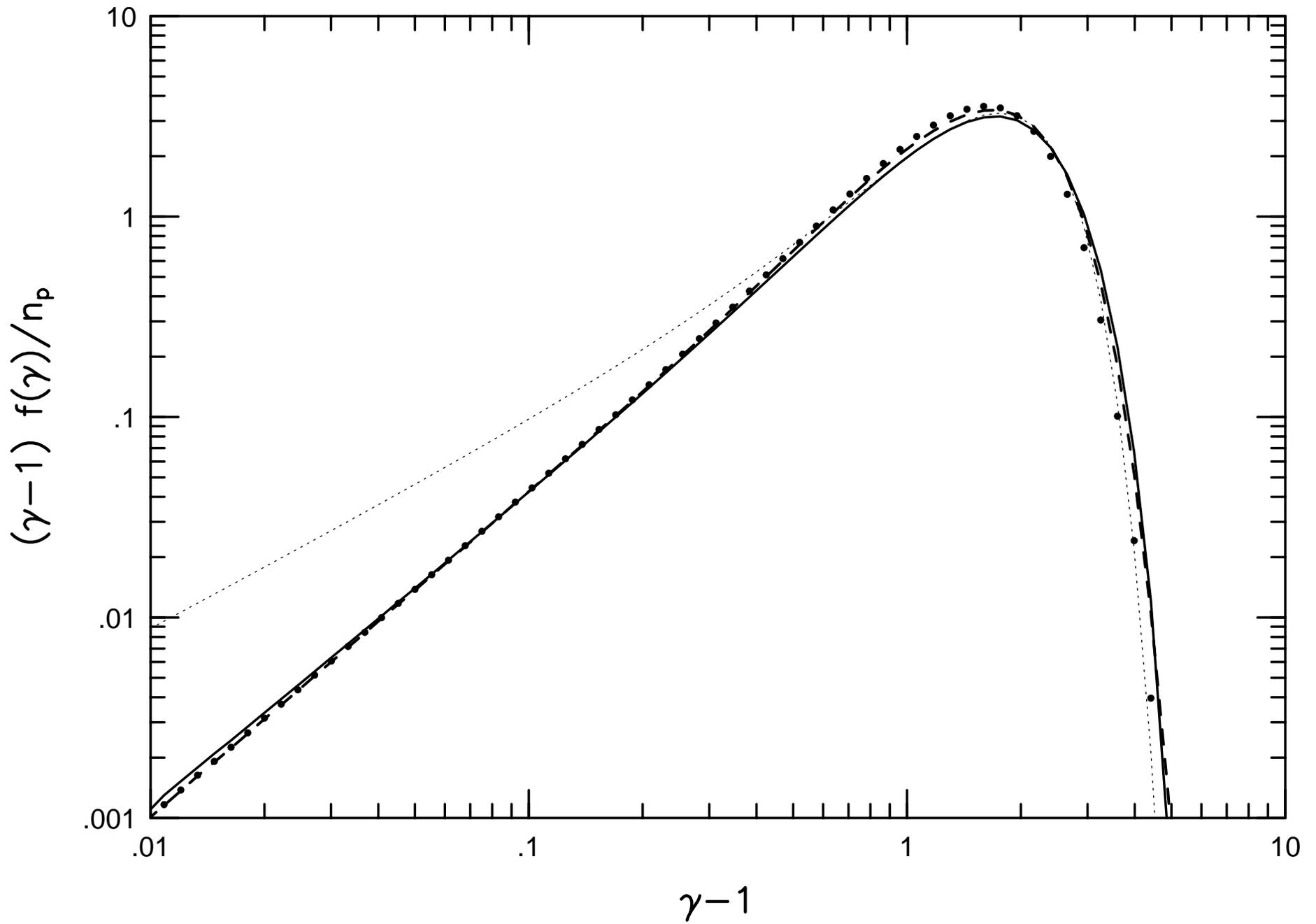
\noindent {\bf Figure 14.} Various fits to the exact electron distribution
(shown with the solid curve) for the test presented in Fig. 12.
The dashed curve corresponds to the function
given by Equation (\vuola) with all the FP coefficients included,
and the big dotted curve corresponds to the solution with the 
electron-electron FP coefficients excluded. 
The `faint' dashed curve  shows the fit given by the simple
expression in Equation (\crude). See text for details.

\vfill\eject\null
\vskip 1.5 cm

\noindent {\bf Table 1.} The output and input parameters for the tests
presented in section 6.3 and plotted in Figures 11 \& 12.
\vskip 0.2in
\centerline{
\vbox{\offinterlineskip
\hrule
\def\trule{\noalign{\hrule}}
\def\vd{&\omit&}     
\def\wrule{height6pt\vd\vd\vd\vd\vd\vd\vd\vd\cr}
\halign{&\vrule#&
	\strut\quad#\quad\cr
\wrule
	&Model$^{\rm a}$&&$l_h$&&$l_s$&&$\tau_p$&&$\Theta_b$&&$\tau_T^{\rm b}$&&
$\langle E\rangle^{\rm c}$&&$kT_p^{\rm d}$ (MeV)&\cr
\wrule
\trule
\wrule
	&exact&&420&&420&&0.05&&$10^{-4}$&&0.22&&0.58&&2400&\cr
\wrule
\wrule
	&wide&&420&&420&&0.05&&$10^{-4}$&&0.41&&0.29&&190&\cr
\wrule
\wrule
	&thermal&&420&&420&&0.05&&$10^{-4}$&&0.27&&0.48&&1100&\cr
\wrule
\trule
\wrule
	&exact&&8.4&&2.1&&0.02&&$3\times 10^{-5}$&&0.14&&1.46&&440&\cr
\wrule
\wrule
	&wide&&8.4&&2.1&&0.02&&$3\times 10^{-5}$&&0.20&&0.87&&53&\cr
\wrule
\wrule
	&thermal&&8.4&&2.1&&0.02&&$3\times 10^{-5}$&&0.17&&1.27&&240&\cr
\wrule
}\hrule}}

\vskip 1 cm

\noindent\null$^{\rm a}$
The assumed or exactly determined shape of the electron distribution
(see \S\ 6.3).  The next four columns give the input parameters for
the various models: hard and soft compactness 
(Eq. 50 \& 54, respectively); $\tau_p$, the `proton optical depth';
the injected blackbody photon temperature, $\Theta_b\equiv kT_b/m_e c^2$.

\noindent\null$^{\rm b}$ Thomson optical depth of the plasma, $\tau_T\equiv
\tau_p(1+2 z)$, where $z$ is the ratio of positron number density to that 
of the protons.

\noindent\null$^{\rm c}$ Average electron energy.

\noindent\null$^{\rm d}$ Proton `temperature', defined in \S\ 6.4.

\noindent Note that differences in the equilibrium optical depth, and
the electron and proton `temperatures' occur even when the radiation spectra 
are almost indistinguishable.

\vfill\eject\null
\centerline{\bf References}
\medskip
\ref Baring, M. 1987, \mnras,  228, 695
\ref Boettcher, M. 1996, private communication.
\ref Cameron, R.A. et al. 1995, Proc. of the Compton Symp., ed. N.
Gehrels, in press
\ref Chandrasekhar, 1942, S. Principles of Stellar Dynamics (Chicago: UC Press),
p. 89
\ref Coppi, P.S. \& Blandford, R.D., 1990, \mnras,  245, 453-469
\ref Dermer, C. D., 1985, \apj,  295, 28
\ref Dermer, C.D., 1984, \apj,  280, 328
\ref Dermer, C. D. \& Liang, E.P., 1989, \apj,  339, 512-528 (DL89)
\ref van Dijk, R. et al. 1995, A\&A, 296, L33
\ref Fabian, A.C. 1994, \apj Supplement Series, 92, p. 555
\ref Guilbert, P.W., Fabian, A.C. \& Stepney, S. 1982, MNRAS, 199, 19p
\ref Ghisellini, G., Guilbert, P. \& Svensson, R. 1988, \apj, 334, L5-L8
\ref Ghisellini, G., Haardt, F. \& Fabian, A.C. 1993, MNRAS, 263, L9 (GHF)
\ref Guilbert, P. W., and Stepney, S. 1985, \mnras,  212, 523
\ref Guilbert, P.W., Fabian, A.C. \& Stepney, S. 1982, MNRAS, 199P, 19
\ref Haardt, F. \& Maraschi, L. 1991, ApJ, 380, L51
\ref Haardt, F. \& Maraschi, L. 1993, ApJ, 413, 507
\ref Haardt, F., Maraschi, L. \& Ghisellini, G. 1994, ApJ, 432, L95
\ref Haardt, F., Maraschi, L. \& Ghisellini, G. 1996, ApJ, submitted.
\ref Jauch, J.M., and Rohrlich, F., 1980, The Theory of Photons and 
Electrons (New York: Springer)
\ref Jones, F.C., 1968, \physrev {\it D}, 167, 1159
\ref Johnson, W.N., et al. 1993, A\&AS, 97, 21
\ref Kompaneets, A.S. 1957, Sov. Phys. JETP, 4, 730
\ref Kusunose, M. 1987, \apj, 321, 186
\ref Landau, L. \& Lifshitz, E., Physical Kinetics, volume 10 of
Course of Theoretical Physics, Pergamon Press, 1981.
\ref Li, Hui, Kusunose, M. \& Liang, E.P. 1996, ApJ Letters, 460, L29
\ref Liang, E.P. 1979, \apj, 234, 1105
\ref Lightman, A.P. \& Zdziarski, A.A. 1987, \apj, 319, 643
\ref Ling, J.C., Mahoney, W.A., Wheaton, Wm. A. \& Jacobson, A.S. 
1987, ApJ, 321, L117
\ref Madejski, G. et al. 1995, \apj, submitted
\ref Maisack, M. et al. 1993, \apj, 407, L67
\ref McConnell, M., et al. 1994, ApJ, 424, 933
\ref Melia, F. \& Misra, R. 1993, \apj, 411, 797
\ref Misra, R. \& Melia, F. 1996, ApJ, 467, 405
\ref Nagirner, D.I., \& Poutanen, J. 1994, Astrophys. Space Phys.
Reviews, Vol. 9, 1-83 
\ref Nayakshin \& Melia, ``Physical Constraint on Active Regions in Seyfert
Galaxies, in preparation'', 1997a
\ref Nayakshin \& Melia, ``Magnetic flares and observed $\tau_T = 1$ in
Seyfert Galaxies'', submitted to ApJL, 1997b
\ref Poutanen, J., \& Svensson, R. 1996, \apj, in press
\ref Press, W.H., Flannery, B.P., Teukolsky, S.A. \& Vetlerling, W.T.
1986, Numerical Recipes (New York: Cambridge)
\ref Rybicki, G. B. \& Lightman, A.P., 1979, Radiative Processes in
Astrophysics, Wiley-Interscience Publication
\ref Shapiro, S.L., Lightman, A.P. \& Eardley, D.M. 1976, \apj, 204, 187
\ref Stepney, S., 1983, \mnras, 202, 467
\ref Sunyaev, R.A. \& Titarchuk, L. 1980, A\&A, 86, 121
\ref Sunyaev, R.A. et al. 1991, Sov. Astron. Letters, 17, 409 
\ref Svensson, R. 1982, \apj, 258, 335
\ref Svensson, R. 1990, in Proc. NATO Advanced Research Workshop on
Physical Processes in Hot Cosmic Plasmas, ed. W. Brinkmann (Dordrecht:
Kluwer), 357
\ref Svensson, R., 1994, ApJS, 92, 585
\ref Svensson, R., 1996,  A\&A, in press.
\ref Titarchuk, L. 1994, ApJ, 434, 570
\ref Titarchuk, L. \& Mastichiadis, A. 1994, ApJ, 433, L33
\ref Zdziarski, A.A., Coppi, P.S. \& Lamb, D.Q. 1990, \apj, 357, 149 
\ref Zdziarski, A.A. et al. 1994, MNRAS, 269, L55
\ref Zdziarski, A.A., Johnson, W.N., Done, C., Smith, D. \&
McNaron-Brown, K. 1995, ApJ, 438, L63
\vfill\eject

\centerline{\bf Appendix}
\medskip
\centerline{\sl A.1 The Finite Differencing scheme}
\medskip
Finding a numerical scheme to solve Equation (\fp) that is stable and preserves
the essential physics (i.e., one that conserves the particle number and energy,
and that leads to a perfect Maxwellian in equilibrium) can be a non-trivial problem.
We shall mention that the need to conserve the number and energy 
is not simply an aesthetic goal, but comes rather from a need to
address practical problems involving pairs, photons and protons (see \S\ 6). 
When the escape rate of particles becomes small (i.e., when the optical 
depth is large), we find severe problems with all differencing schemes 
that do not conserve the energy and number \lq\lq exactly\rq\rq.
In particular, unphysical pair runaways may ensue.
We shall therefore present the energy and
particle conserving scheme in detail.  The scheme we have settled on is
$$\eqnam{\scheme}
f_k^{n+1} = f_k^n -
\Delta t \Bigl ({a_{k+1}^n f_{k+1}^n - a_{k-1}^n f_{k-1}^n
\over \Delta x_{k-1}+\Delta x_{k}}\Bigr )
+{s\over 2}\, \Bigl [ \alpha D_{k+1}^n f_{k+1}^n
- 2  D_{k}^n f_{k}^n +\rho  D_{k-1}^n f_{k-1}^n \Bigr ],
\eqno(A1)
$$
where
$$
s \equiv {2 \Delta t \over \alpha\,\Delta x_k\,
(\Delta x_{k-1} + \Delta x_k)}\;.
\eqno(A2)
$$
Here $a(E)$ and $D(E)$ are the energy exchange and diffusion FP 
coefficients, respectively, which are given as integrals over the 
distribution function $f$ itself.

The index $k$ refers to the energy ($x_k$ is a point in energy space), and $n$ denotes
the time.  The parameters $\alpha$ and $\rho$ have been 
introduced to make sure the scheme 
preserves the particle normalization and energy.  In our calculations, we have chosen a
logarithmic energy spacing, and one can show in this case that in order to preserve these
aspects, one must choose
$$\eqnam{\param}
\alpha = {2\over 1+q}, \doublespace\,\,\,\,\,\,\,\,\,\,\,
\,\,\,\,\,\,\,\,
\rho = {2 q \over 1+q}\;,
\eqno(A3)
$$
where $q=x_{k+1}/x_k$. To demonstrate that this must be the 
case, we need only consider a
numerical realization of the integrals $\int dE\, [f^n(E)-f^{n+1}(E)]$
and $\int dE\, E [f^n(E)-f^{n+1}(E)]$ for the evolving distribution,
and require these to give zero. 

Because Equation (\fp) is a flux equation, a sensible 
constraint to impose at the boundary
is a mirror condition at both the low and high energy ends.  This means modifying 
the differencing scheme (Eq. A1) for a few boundary points. We explicitly set 
the value of the distribution function in the first and last points to zero. 
That means that the two extreme points in the actual distribution function should
be far enough from the maximum of the distribution that neglecting the 
(small) number
of particles in the boundary bins does not introduce a noticeable error. 
Requiring the total number of electrons to not change with time, one
obtains certain conditions on the differencing scheme at the neighborhood
of the boundaries. These conditions lead us to choose the differencing scheme
for the second point on the energy axis to be 
$$\eqnam{\secp}
f_2^{n+1} = f_2^n -
\Delta t\, \Bigl ({a_{3}^n f_{3}^n + a_{2}^n f_{2}^n
\over \Delta x_{1}+\Delta x_{2}}\Bigr )
+{s\over 2}\, \Bigl [ \alpha D_{3}^n f_{3}^n
-   D_{2}^n f_{2}^n \Bigr ],
\eqno(A4)
$$
and for the next to last point, 
$$\eqnam{\lastp}
f_M^{n+1} = f_M^n +
\Delta t\, \Bigl ({a_{M}^n f_{M}^n + a_{M-1}^n f_{M-1}^n
\over \Delta x_{M-1}+\Delta x_{M}}\Bigr )
+{s\over 2}\, \rho\;\Bigl [ -   D_{M}^n f_{M}^n 
+  D_{M-1}^n f_{M-1}^n 
\Bigr ],
\eqno(A5)
$$
where $M= N-1$, and $N$ is the total number of points in energy
space. For the third point ($x_3$), the only change needed to be made is to 
set the coefficient $\rho$ in Equation (A1) to $\rho = q$.

Because Equations (A4) and (A5) employ only two boundary points
instead of three as any other $k \neq 2,M\,$, it is not possible to make
the scheme conserve particle number {\it and} energy simultaneously
for these two bins.
We need two parameters to conserve these integrals, while we have only one 
parameter available in Equations (A4) and (A5) (i.e., the ratio $\alpha/\rho$).
Physically, by bouncing particles scattering into the first bin back
to the second and third bins ($f_1(t)\equiv 0 $ at all times), we
conserve the number of particles, but {\it change} the energy of 
these particles to make them stay inside this energy
range (i.e., inside the region $x_2\le \gamma -1\le
x_{N-1}$). This process therefore does not conserve energy,
and we must subtract this excess energy from the total
energy change during each iteration. 
Practically, this is done by introducing a small correction 
to the electron energy exchange coefficient, which allows us to conserve energy
up to double precision accuracy during the entire particle thermalization time.

Note that the mirror boundary conditions are perfectly physical and are
applicable to any electron distribution function with large or small 
deviations from a Maxwellian. 
Indeed, if there were a physically substantial
flow through the low energy boundary, that would mean that there is either
a source or sink of particles with $E\ll \Theta_e$, 
where $\Theta_e $ is the dimensionless electron 
temperature, $\Theta_e \equiv k T_e/m_e c^2$. While this situation is 
mathematically possible, it is not plausible physically (if there is an
influx of cold electrons from outside into the system, one can always choose
the low energy boundary such that it includes these electrons). 
On the high energy end, any electron acceleration mechanism can 
only accelerate electrons to some  maximum energy (specific for this
mechanism), after which electron cooling overcomes the heating. Therefore,
if one chooses the high energy boundary much higher than this maximum energy,
then there can be no flux through the boundary, since the electrons get 
\lq turned back\rq$\,$ to lower energies before they reach the boundary.

The ultimate test of the given scheme is to see how close the equilibrium
distribution function approaches a Maxwellian (when one includes only
the lepton-lepton FP part of the full lepton Boltzmann equation).
We have run a number of tests
(described in the next few sections) and find excellent agreement between 
the equilibrium distribution and a Maxwellian (with a relative
deviation of less than $2\%$).
\medskip
\centerline{\sl A.2 Numerical Stability Of The Scheme}
\medskip

Equation (A1) is a non-linear integro-differential equation, for which
strictly speaking, the usual Von Neumann analysis should not
be used to test its stability.
Fortunately, the fact that the FP coefficients are integrals over
the distribution function makes the stability analysis
essentially linear. 
Conducting the stability analysis 
of our scheme as if $a(E,t)$ and $D(E,t)$ are fixed functions,
or at most, slowly varying functions compared to $f(E,t)$,
and using the usual Von-Neuman approach (e.g., Press et al. 1986), 
we arrive at the approximate stability criterion:
$$\eqnam{\stabil}
\max\,\Bigl \{ \Bigl(1-
{D_k^n \Delta t\over (\Delta x_k)^2} \Bigr )^2,
 \bigl( {a_k \Delta t\over \Delta x_k}\bigr )^2
\Bigr \}< 1\;.
\eqno( A6)
$$
Equation (A6) is a necessary condition for the scheme (Eq. A1) to be 
numerically stable.  While this condition is only approximate, 
we find that by choosing a time step roughly equal to or smaller
than that which satisfies Equation (A6), our differencing 
scheme is always stable. 

For applications where `exact' energy conservation is not
an issue (e.g., in systems with a low optical depth and a high 
compactness, such that the energy is radiated efficiently from the 
system), we were able to develop a half implicit-half explicit 
scheme that is much faster than the explicit scheme (see the next section).

\bigskip
\centerline{\sl A.3 The Numerical Procedure}
\medskip
We follow the procedure outlined in \S A.1 above, except that
we here include the additional terms corresponding to Compton 
interactions and proton heating.
The particle distributions are evolved more often than the photon 
distribution, i.e., the photon distribution is kept fixed during a
time $\Delta t_M=M\Delta t$ while the particles proceed through
their evolution a number $M$ ($\gg 1$) of $\Delta t$ steps.
This approach is necessary because otherwise in each
time advancement, one needs to re-evaluate the photon annihilation 
cross section and the Compton scattering matrix, among others. 
This increases the computation time considerably. 
The value of $\Delta t$ is dictated by the stability criterion 
given in Equation (A6). 
It turns out that this condition is much stricter than any other 
criterion for $\Delta t$ based, e.g., on the requirement that 
distribution functions change very little during one time advancement
(i.e. $|\Delta f(E)| \ll |f(E)|$).  The latter should be
used for determining the time step when evolving the photon distribution 
since there is no numerical instability problem analogous to 
Equation (\photeq).  We are therefore justified in using 
$\Delta t_M \gg \Delta t$.

The computation time depends considerably on whether one uses  explicit
or implicit schemes and on the Thomson optical depth of the system.
For a Thomson optical depth $\tau_T\sim 1$, the explicit scheme takes 
a few days on a Pentium PC running Linux 
(for 70 bins in the electron energy space and 70 bins for photons), whereas 
an implicit scheme takes from 6 hours to a day. Obviously, such a long
computing time makes it difficult to use the described methods 
in practice, for example to fit observations. 

For most of the applications, however, a time-independent code should
suffice. We have been able to write and use such a code successfully.
Using the approximate solution of the full electron equation given by
Equation (\vuola), and assuming that the positron distribution is
same as that of the electrons, we were able to reduce computing time
to $2-5$ minutes. In essence, the code is not any slower than a 
corresponding code that assumes that particles are thermal and takes into account 
the same interactions and follows radiation transport in the same manner. 
Thus, the time-independent code is highly efficient and might easily be used
in X-ray fitting packages, such as XSPEC. This code and some results are to
be published elsewhere (the results presented in the current paper were obtained 
using the
full time-dependent implicit code). The authors can provide their code for use
by anybody who feels a need to use exact particle distributions. 

To verify that our code was correctly resolving the particle distribution
in energy space, we also varied the number of energy points.
A comparison of our results from a test with 100 energy bins (for both
electrons and photons) with those from a simulation with 70,
showed that the relative differences in the distribution, 
spectrum, temperature and the number of positrons were $\simlt 1\%$.

\bigskip
\centerline{\sl A.4 Large Angle Scatterings}
\medskip

As already mentioned in \S 3, it has been known since the early work of
Chandrasekhar and Landau that large angle scattering events are not important
for Coulomb interactions in a gas, since the rate of energy exchange due to 
these is much smaller than that due to small angle
collisions. This point is worth discussing further here, since 
it appears to be the basis for the difference between our approach
and that of Dermer \& Liang 1989. 
Large angle scatterings correspond to small impact parameters. We can estimate
the latter for large angle scatterings $r_l$ by writing
$$\eqnam{\rlarge}
{e^2\over r_l}\simgt |\Delta E| m_e c^2
\eqno(A7)
$$
(recalling that $E\equiv \gamma - 1$ is dimensionless).
The cross section is then $\sigma_l(\Delta E) \simlt \pi r_l^2$, and the energy 
exchange rate is 
$$\eqnam{\elar}
a_l(E) \sim \sigma_l(\Delta E) |\Delta E| n_e \beta c \simlt 
\pi\, {e^4\over m_e^2 c^4}\, n_e c \beta/|\Delta E|\;.
\eqno(A8)
$$
Let us consider the high energy portion of the electron distribution, i.e.,
$E\gg \Theta$. Large angle scatterings correspond to a large change in
the electron energy $\Delta E\sim E$ (Equation [\echange]).
Therefore, we can put $\Delta E =  E$ in the above equation, and 
re-write it as 
$$\eqnam{\elarg}
a_l(E) \sim {3\over 8}\, {1\over t_T}\, {\beta\over \gamma - 1}\;.
\eqno(A9)
$$
Equation (\climit) can be used to estimate the small angle energy exchange
rate
$$\eqnam{\asmal}
a_s(E) \sim 3/2\, {\ln \Lambda \over t_T} \,{1\over \beta
\langle \gamma\rangle }\;,
\eqno(A10)
$$
where $\langle \gamma\rangle$ is the average thermal $\gamma$-factor.
The ratio of the large angle energy exchange rate to that of
small angle scatterings is thus 
$$\eqnam{\arat}
{a_l(E)\over a_s(E)}\sim {\beta^2 \langle \gamma\rangle \over
\gamma-1} \,{1\over 4 \ln \Lambda} \ll 1\;,
\eqno(A11)
$$
as long as $\ln \Lambda \gg 1$. Let us now assume $E\ll \Theta$. In this case
we should use the average thermal $\beta$ and $\gamma$ in Equation (A9),
since the energy exchange can be as large as $\sim \Theta$, and $\beta$ in
equation (A8) is really the relative velocity of the two colliding
particles, which is of the same order as the thermal velocity. The ratio
of the scattering rates becomes
$$\eqnam{\arata}
{a_l(E)\over a_s(E)}\sim {\langle \beta\rangle^2 \gamma \over 
\langle \gamma\rangle - 1}\,{1\over 4 \ln \Lambda} < {1\over 4 \ln \Lambda}\;.
\ll 1
\eqno(A12)
$$
A similar estimate holds for energies $E\sim \langle E\rangle$, except
for the narrow region where the small angle energy-exchange coefficient 
goes to zero (Fig. [3]). 

Therefore, it is evident that one may neglect
the large angle scatterings in the full kinetic equation that describes
Coulomb collisions. Moreover, we argue that it is even necessary to do so
in the calculation of the FP coefficients, since otherwise the approximation
made for small scattering angles becomes invalid for large angles
and can lead to wrong conclusions. Physically, one cannot decompose the 
electron distribution function in the original Boltzmann kinetic equation,
since the large angle scatterings do not constitute a diffusion process.
For example, the use of the diffusion
coefficient as computed by Dermer \& Liang (1989) can lead to a power-law 
electron distribution on the high energy end instead of a Maxwellian distribution,
if other interactions, such as
inverse Compton cooling are not important. The large angle
scattering terms, which according to the above discussion should contribute very
little, in fact start to dominate the diffusion coefficient. The energy exchange 
coefficient, on the other hand, is unaffected by the inclusion of these
terms since (by coincidence) averaging over $\phi^*$ cancels out the large 
positive and negative $\Delta E$ in Equation (\echange).
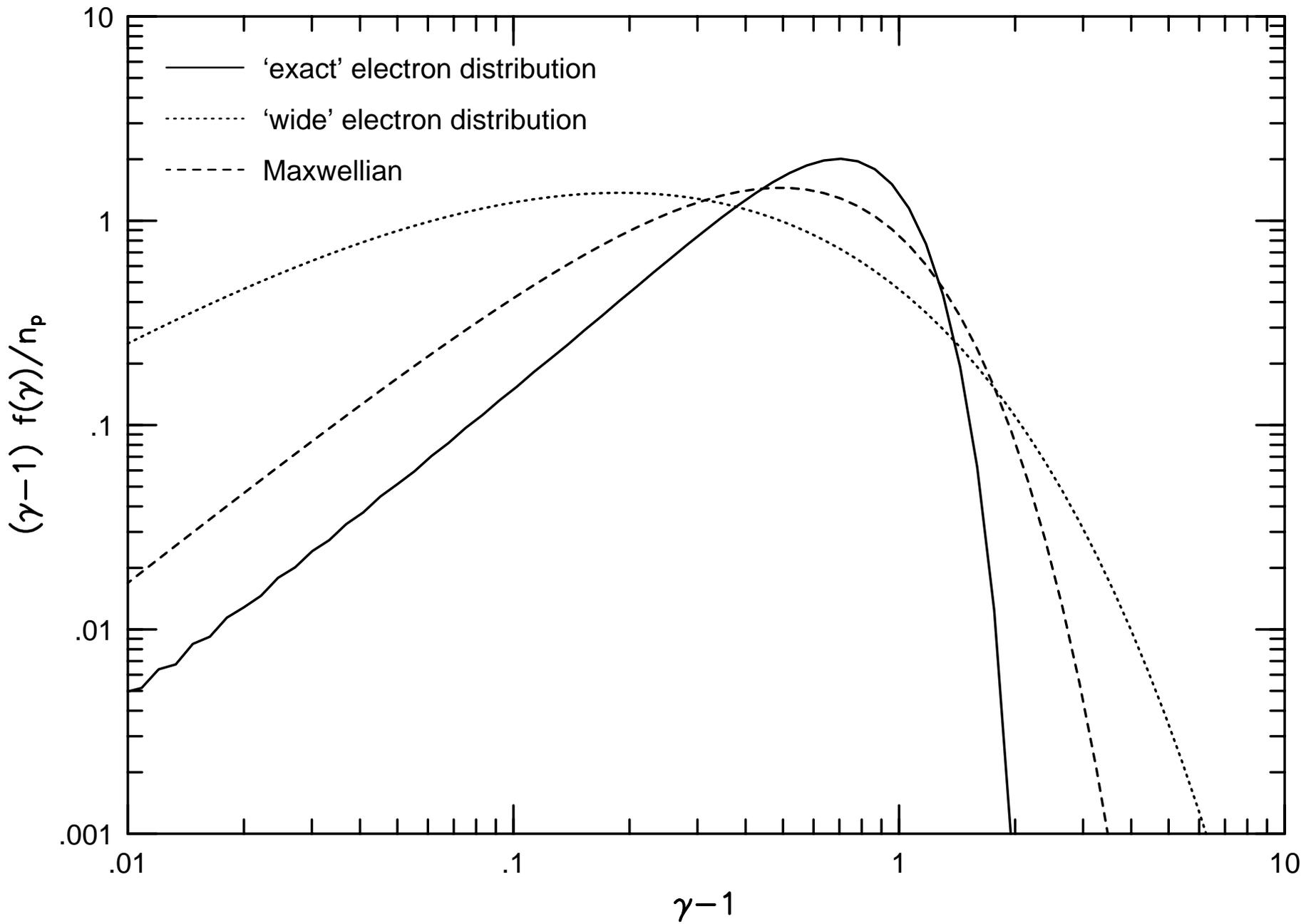
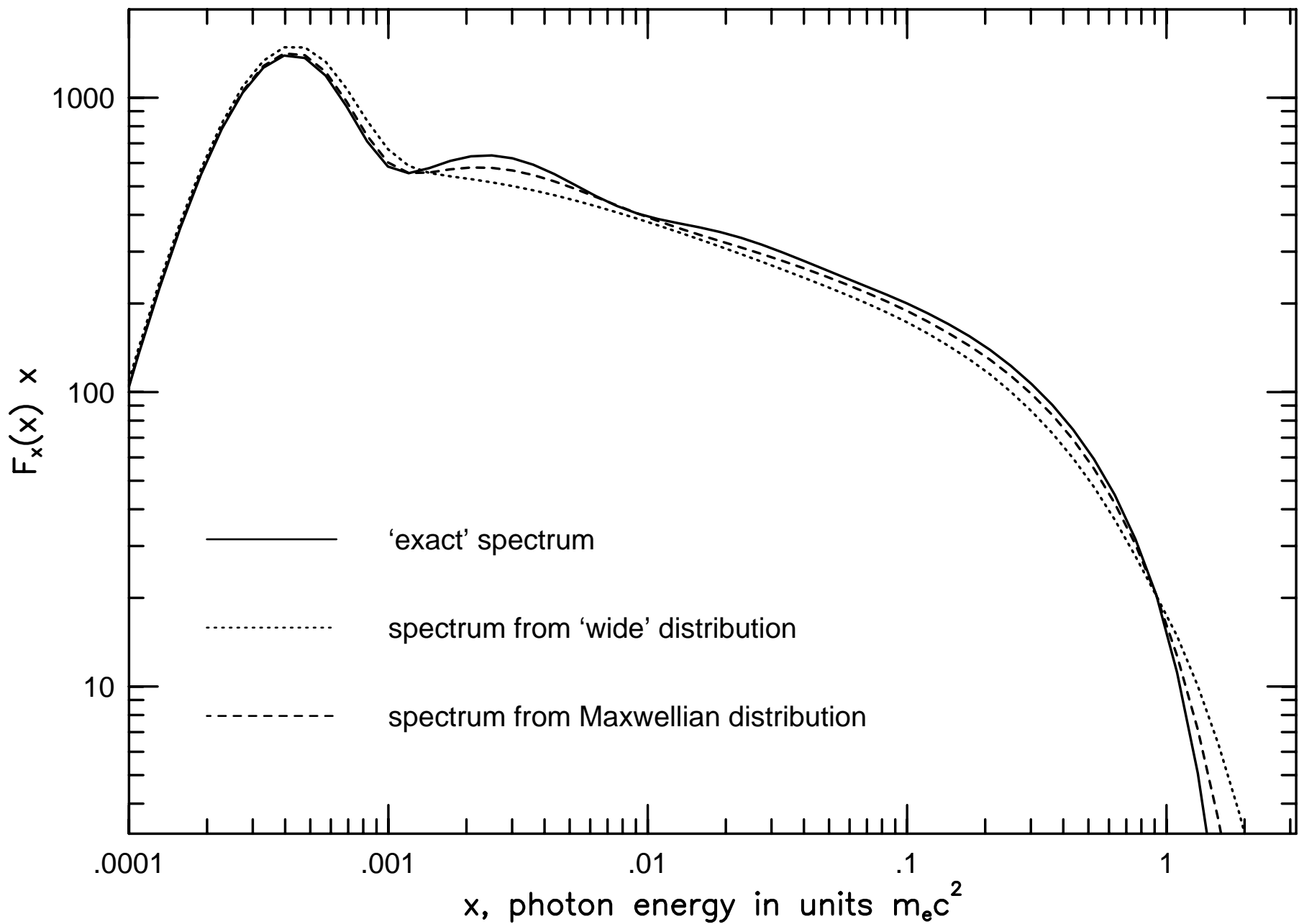
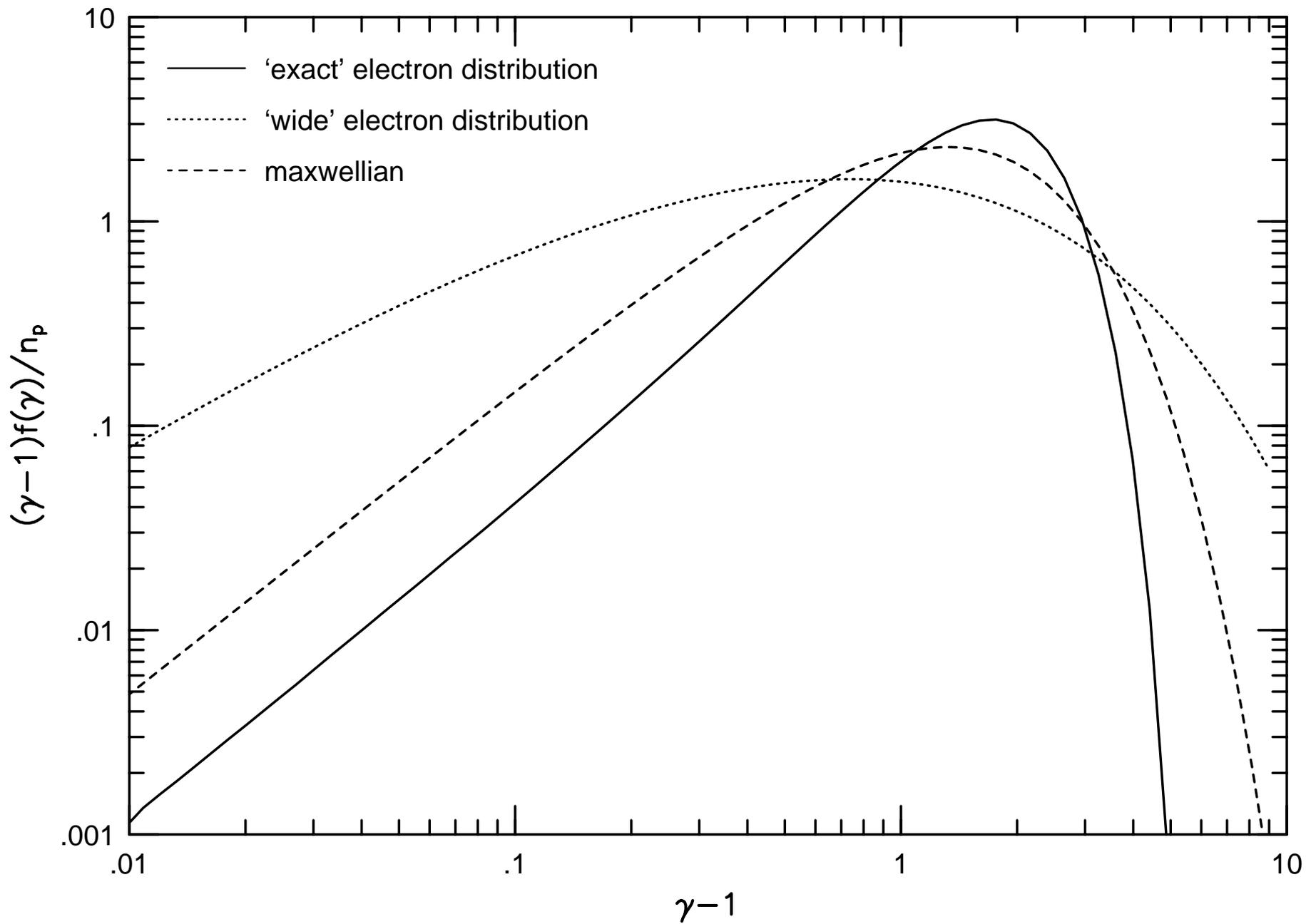
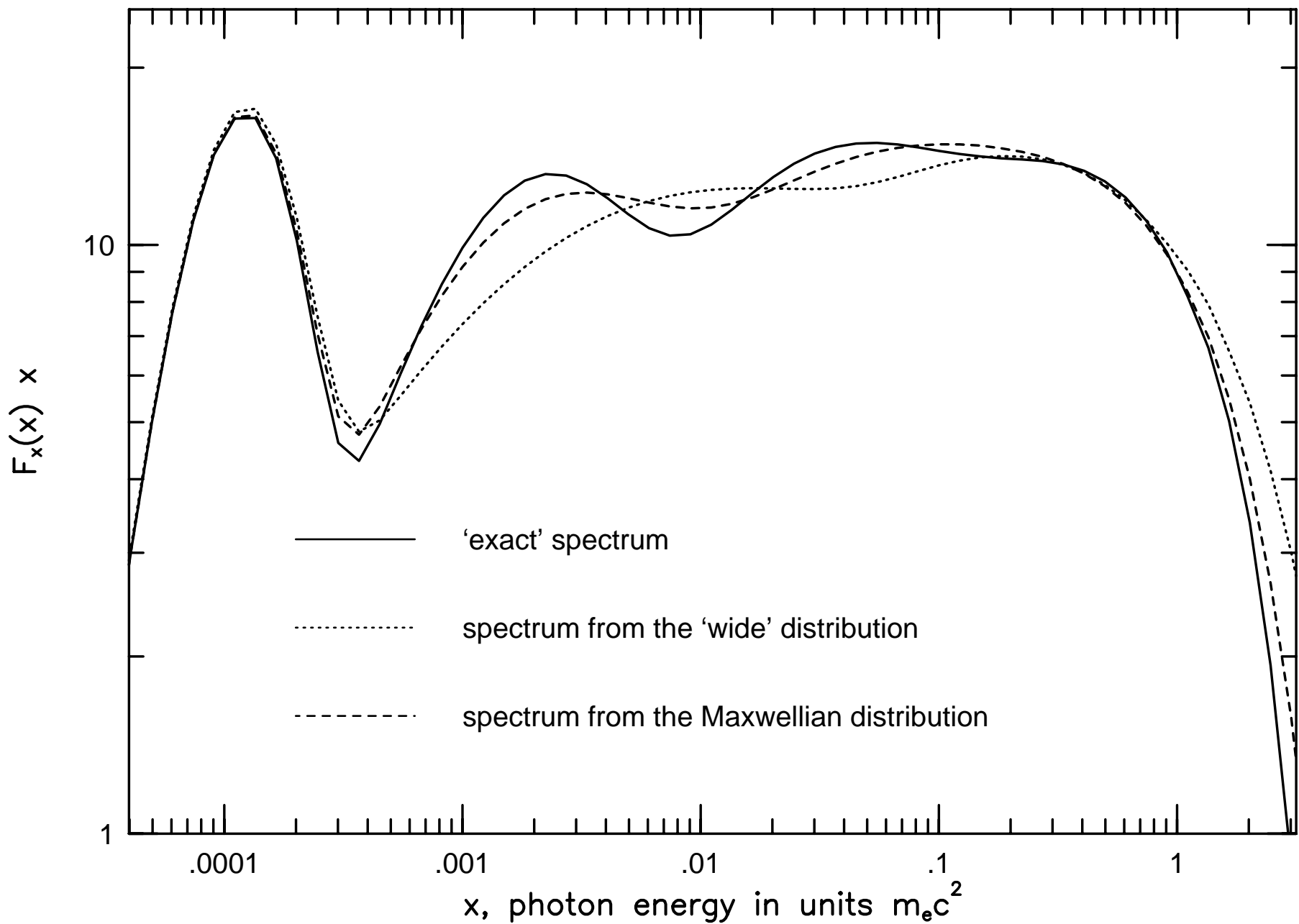
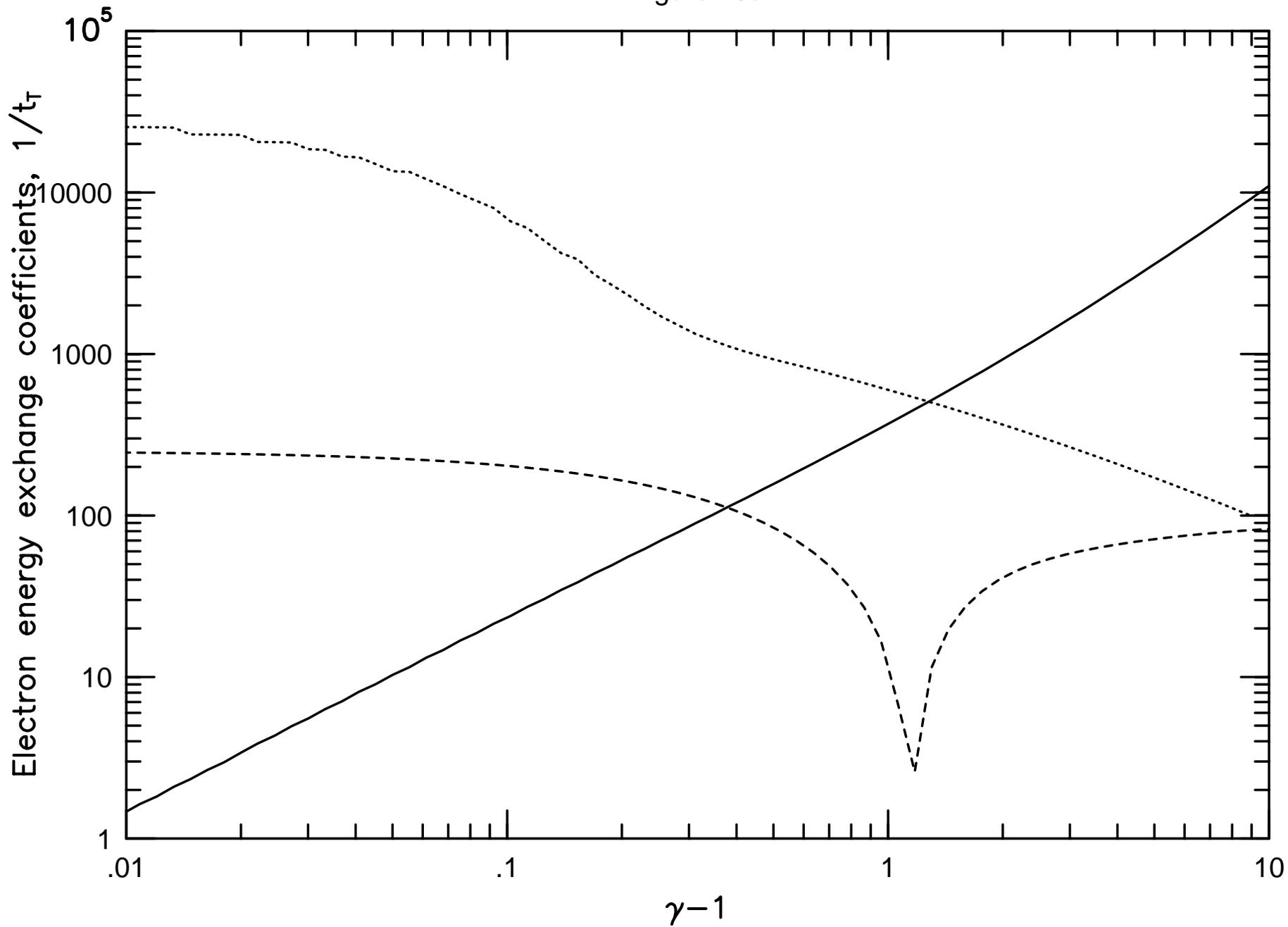
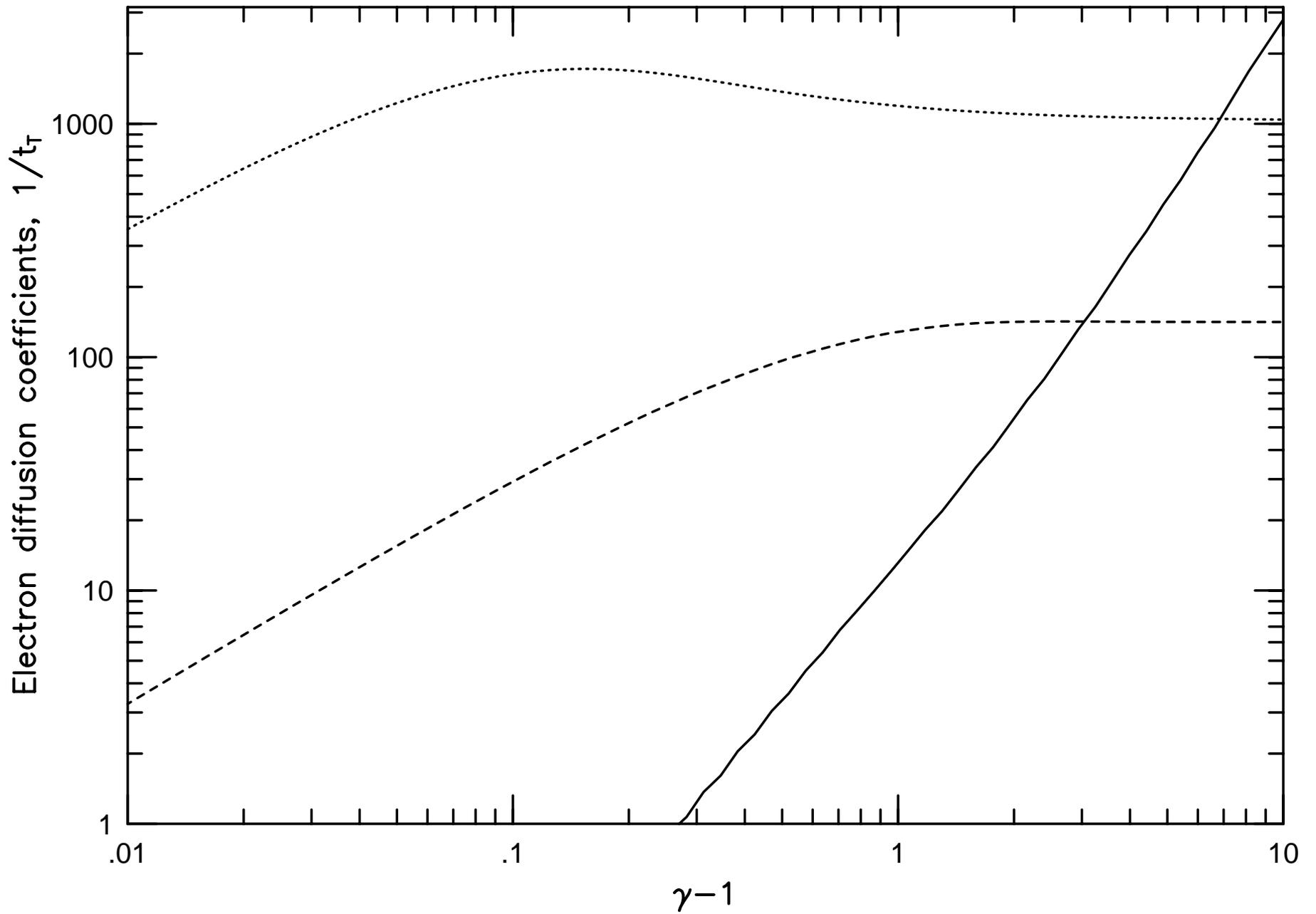
\end